\documentclass[sn-mathphys,Numbered]{sn-jnl}

\usepackage{graphicx}%
\usepackage{multirow}%
\usepackage{amsmath,amssymb,amsfonts}%
\usepackage{amsthm}%
\usepackage{mathrsfs}%
\usepackage[title]{appendix}%
\usepackage{xcolor}%
\usepackage{textcomp}%
\usepackage{manyfoot}%
\usepackage{booktabs}%
\usepackage{algorithm}%
\usepackage{algorithmicx}%
\usepackage{algpseudocode}%
\usepackage{listings}%
\usepackage{subfig} 
\usepackage{comment} 
\usepackage{lineno}

\theoremstyle{thmstyleone}%
\theoremstyle{thmstyletwo}%
\theoremstyle{thmstylethree}%

\begin{document}

\title[Article Title]{Preliminary results on the long term operation of RPCs with eco-friendly gas mixtures under irradiation at the CERN Gamma Irradiation Facility}



\author*[1]{\small \textbf{The RPC ECOgas@GIF++ Collaboration: }L. Quaglia} \email{\small luca.quaglia@cern.ch}
\author*[2,3]{\small D. Ramos} \email{\small dayron.ramos.lopez@cern.ch}
\author[6,3]{\small M. Abbrescia}
\author[21]{\small G. Aielli}
\author[3,8]{\small R. Aly}
\author[17]{\small M. C. Arena}
\author[11]{\small M. Barroso}
\author[4]{\small L. Benussi}
\author[4]{\small S. Bianco}
\author[7]{\small D. Boscherini}
\author[16]{\small F. Bordon}
\author[7]{\small A. Bruni}
\author[18]{\small S. Buontempo}
\author[16]{\small M. Busato}
\author[21]{\small P. Camarri}
\author[13]{\small R. Cardarelli}
\author[3]{\small L. Congedo}
\author[11]{\small D. De Jesus Damiao}
\author[6,3]{\small M. De Serio}
\author[21]{\small A. Di Ciacco}
\author[21]{\small L. Di Stante}
\author[14]{\small P. Dupieux}
\author[19]{\small J. Eysermans}
\author[12,1]{\small A. Ferretti}
\author[6,3]{\small G. Galati}
\author[12,1]{\small M. Gagliardi}
\author[12,1]{\small S. Garetti}
\author[16]{\small R. Guida}
\author[2,3]{\small G. Iaselli}
\author[14]{\small B. Joly}
\author[24]{\small S.A. Juks}
\author[23]{\small K.S. Lee}
\author[13]{\small B. Liberti} 
\author[22]{\small D. Lucero Ramirez}
\author[16]{\small B. Mandelli}
\author[14]{\small S.P. Manen}
\author[7]{\small L. Massa}
\author[3]{\small A. Pastore}
\author[13]{\small E. Pastori}
\author[4]{\small D. Piccolo}
\author[13]{\small L. Pizzimento}
\author[7]{\small A. Polini}
\author[13]{\small G. Proto}
\author[2,3]{\small G. Pugliese}
\author[16]{\small G. Rigoletti}
\author[13]{\small A. Rocchi}
\author[7]{\small M. Romano}
\author[10]{\small A. Samalan}
\author[9]{\small P. Salvini}
\author[21]{\small R. Santonico}
\author[5]{\small G. Saviano}
\author[13]{\small M. Sessa}
\author[6,3]{\small S. Simone}
\author[12,1]{\small L. Terlizzi}
\author[10,20]{\small M. Tytgat}
\author[12,1]{\small E. Vercellin}
\author[15]{\small M. Verzeroli}
\author[22]{\small N. Zaganidis}

\affil[\footnotesize 1]{\footnotesize INFN Sezione di Torino, Via P. Giuria 1, 10126 Torino, Italy}
\affil[\footnotesize 2]{\footnotesize Politecnico di Bari, Dipartimento Interateneo di Fisica, via Amendola 173, 70125 Bari, Italy}
\affil[\footnotesize 3]{\footnotesize INFN Sezione di Bari, Via E. Orabona 4, 70125 Bari, Italy}
\affil[\footnotesize 4]{\footnotesize INFN - Laboratori Nazionali di Frascati, Via Enrico Fermi 54, 00044 Frascati (Roma), Italy}
\affil[\footnotesize 5]{\footnotesize Sapienza Università di Roma, Dipartimento di Ingegneria Chimica Materiali Ambiente, Piazzale Aldo Moro 5, 00185 Roma, Italy}
\affil[\footnotesize 6]{\footnotesize Università degli studi di Bari, Dipartimento Interateneo di Fisica, via Amendola 173, 70125 Bari, Italy}
\affil[\footnotesize 7]{\footnotesize INFN Sezione di Bologna, Viale C. Berti Pichat 4/2, 40127 Bologna, Italy}
\affil[\footnotesize 8]{\footnotesize Helwan University, Helwan Sharkeya, Helwan, Cairo Governorate 4037120, Egypt}
\affil[\footnotesize 9]{\footnotesize INFN Sezione di Pavia, Via A. Bassi 6, 27100 Pavia, Italy}
\affil[\footnotesize 10]{\footnotesize Ghent University, Dept. of Physics and Astronomy, Proeftuinstraat 86, B-9000 Ghent, Belgium}
\affil[\footnotesize 11]{\footnotesize Universidade do Estado do Rio de Janeiro, R. São Francisco Xavier, 524 - Maracanã, Rio de Janeiro - RJ, 20550-013, Brasil}
\affil[\footnotesize 12]{\footnotesize Università degli studi di Torino, Dipartimento di Fisica, Via P. Giuria 1, 10126 Torino, Italy}
\affil[\footnotesize 13]{\footnotesize INFN Sezione di Roma Tor Vergata, Via della Ricerca Scientifica 1, 00133 Roma, Italy}
\affil[\footnotesize 14]{\footnotesize Clermont Université, Université Blaise Pascal, CNRS/IN2P3, Laboratoire de Physique Corpusculaire, BP 10448, F-63000 Clermont-Ferrand, France}
\affil[\footnotesize 15]{\footnotesize Université Claude Bernard Lyon I, 43 Bd du 11 Novembre 1918, 69100 Villeurbanne, France}
\affil[\footnotesize 16]{\footnotesize CERN, Espl. des Particules 1, 1211 Meyrin, Svizzera}
\affil[\footnotesize 17]{\footnotesize Università degli studi di Pavia, Corso Strada Nuova 65, 27100 Pavia, Italy}
\affil[\footnotesize 18]{\footnotesize INFN Sezione di Napoli, Complesso Universitario di Mone S. Angelo ed. 6, Via Cintia, 80126 Napoli, Italy}
\affil[\footnotesize 19]{\footnotesize Massachusetts Institute of Technology, 77 Massachusetts Ave, Cambridge, MA 02139, USA}
\affil[\footnotesize 20]{\footnotesize Vrije Universiteit Brussel (VUB-ELEM), Dept. of Physics, Pleinlaan 2, 1050 Brussels, Belgium}
\affil[\footnotesize 21]{\footnotesize Università degli studi di Roma Tor Vergata, Dipartimento di Fisica, via della Ricerca Scientifica 1, 00133 Roma, Italy }
\affil[\footnotesize 22]{\footnotesize Universidad Iberoamericana, Dept. de Fisica y Matematicas, Mexico City 01210, Mexico.}
\affil[\footnotesize 23]{\footnotesize Korea University, 145 Anam-ro, Seongbuk-gu, Seoul, South Korea}
\affil[\footnotesize 24]{\footnotesize Université Paris-Saclay, 3 rue Joliot Curie, Bâtiment Breguet 91190 Gif-sur-Yvette, France}








\abstract{Since 2019 a collaboration between researchers from various institutes and experiments (i.e. ATLAS, CMS, ALICE, LHCb/SHiP and the CERN EP-DT group), has been operating several RPCs with diverse electronics, gas gap thicknesses and detector layouts at the CERN Gamma Irradiation Facility (GIF++). The studies aim at assessing the performance of RPCs when filled with new eco-friendly gas mixtures in avalanche mode and in view of evaluating possible ageing effects after long high background irradiation periods, e.g. High-Luminosity LHC phase. This challenging research is also part of a task of the European AidaInnova project.

A promising eco-friendly gas identified for RPC operation is the tetrafluoruropropene (C$_{3}$H$_{2}$F$_{4}$, commercially known as HFO-1234ze) that has been studied at the CERN GIF++ in combination with different percentages of CO$_2$. Between the end of 2021 and 2022 several beam tests have been carried out to establish the performance of RPCs operated with such mixtures before starting the irradiation campaign for the ageing study.

Results of these tests for different RPCs layouts and different gas mixtures, under increasing background rates are presented here, together with the preliminary outcome of the detector ageing tests.}




\maketitle

\section{Introduction}
\label{sec:intro}

The European Union has declared in EU regulation 517/2014 \cite{eu-517-2014}, the phase down and limitation of fluorinated greenhouse gases (GHGs) production and usage. Many scientific centers are pushing experimental collaborations to look for possible eco-friendly replacements for the used gas mixtures. CERN, in particular, is committed to reducing its GHG emissions and phase-down actions have been put in place since 2020 \cite{envReport}. Some studies report RPCs as the major contributor to GHG emission from detector systems at the Large Hadron Collider (LHC) during Run 1 and Run 2 \cite{Mandelli2022}.

The RPCs at CERN are mainly operated with mixtures composed of around 5\% isobutane (i-C$_{4}$H$_{10}$), more than 90\% tetrafluoroethane (C$_2$H$_2$F$_4$), and less than 1\% of SF$_6$. The latter two, are both GHGs characterized by a global warming potential (GWP) of 1430 and 22800, respectively while isobutane has a lower GWP equal to 3 \cite{eu-517-2014}.


The search for eco-friendly alternatives plays a fundamental role in the strategies to reduce GHGs emissions and possibly the relative operational costs. In this context, the RPC ECOgas@GIF++ collaboration, which includes RPC physicists from the LHC experiments (CMS, ALICE, ATLAS, LHCb) and the gas group of CERN, is pursuing the use of new gases in order to find a proper eco-friendly gas mixture replacement. Although the currently employed RPC gas mixture contains two different GHGs components, it is quite challenging to find a replacement for both simultaneously. For this reason, the collaboration started to study alternatives to C$_2$H$_2$F$_4$ since it is the main contributor to the mixture GWP. Possible candidates have been identified in the family of the Hydro-Fluoro-Olefins (HFOs), due to their molecular similarity with the C$_2$H$_2$F$_4$ and lower GWP ($\sim$6).

Within this gas family, tetrafluoropropene (C$_3$H$_2$F$_4$), which comes in different isomer forms \cite{isomers}, was of particular interest, mainly because two isomers, the HFO-1234ze and HFO-123yf, are currently used as refrigerants in the industry, making them rather available for purchase. The choice fell on HFO-1234ze due to the mild flammability of the yf isomer \cite{NEEDHAM2017176}, making it unsuitable to be used in the LHC experiments due to safety reasons. In what follows, HFO-1234ze will be referred to as HFO. 

The HFO molecule contains the same number of Hydrogen and Fluorine atoms but one more Carbon with respect to C$_2$H$_2$F$_4$. On the other hand, the full replacement of tetrafluoroethane with this gas leads to an increase of the detector working voltage \cite{increaseWP}. This is not advisable since the currently installed detectors and high voltage systems are not designed to operate at such high voltages, making this simple replacement not possible. A mitigating solution was to dilute the HFO with CO$_2$, reducing the partial pressure of the gas mixture and lowering the operating voltage \cite{Bianchi_2019, prelPiccolo, prelGiorgia, GUIDA2020162073}. In this framework, HFO, being much denser than CO$_2$ will work as the main contributor to the primary ionization. However, the ionizing effects of CO$_2$ have to be into account, especially in those mixtures in which the HFO concentration is lower \cite{sauli, Saviano_2018}.

The efforts of the RPC ECOgas@GIF++ collaboration are two-pronged: on the one hand, various beam test campaigns have been carried out to fully characterize RPC response when operated with various HFO/CO$_{2}$-based gas mixtures. On the other hand, the longevity of the RPCs, when operated with eco-friendly gas mixtures, is studied by performing an aging test under an intense radiation background. During this kind of study, the detectors are exposed to an intense flux of $\gamma$ photons, mimicking the background conditions expected during the High-Luminosity (HL) LHC phase \cite{highLumi} and the stability of their response is studied over time.

    
    

This article summarizes the main results obtained in the above-mentioned beam test campaigns and aging studies. It is organized as follows: section \ref{sec:experimental} contains a brief description of the experimental setup, section \ref{sec:beamTest} reports the results obtained from the 2022 beam test campaign (results from the first RPC ECOgas@GIF++ beam test campaign (in 2021) can be found in \cite{testBeam2021}), section \ref{sec:agingStudies} describes the results obtained so far from the aging campaign. Lastly, section \ref{sec:conclusion} contains a summary of the results, providing also an outlook for possible future developments of these studies.

\section{Experimental setup}\label{sec:experimental}

The studies of the RPC ECOgas@GIF++ collaboration are carried out at the CERN Gamma Irradiation Facility (GIF++). This facility is equipped with a 12.5~TBq $^{\text{137}}$Cs source, which allows the users to perform long-term irradiation studies (aging campaigns) with conditions similar to the ones present during the HL-LHC phase \cite{highLumi}. The GIF++ is built on the H4 secondary SPS beam line and it is traversed by a high-energy (150~GeV) muon beam, produced by the interaction of the SPS proton beam on a number of fixed targets \cite{H4beam}.

The $^{\text{137}}$Cs source is equipped with an Aluminum angular correction filter, used to transform the 1/$r^{\text{2}}$ ($r$ being the distance from the source) dependence of the $\gamma$ flux to a  uniformly distributed flux on the xy plane (perpendicular to the beam), providing uniform irradiation on large area detectors. The irradiation from the $^{\text{137}}$Cs source can be modulated by means of a set of lead attenuation filters which are arranged as a 3$\times$3 array, allowing for a total of 27 possible attenuation values, from 1 (irradiator fully opened) to 46000 (maximum attenuation). The irradiator can also be fully shielded (condition referred to as "source-off" in the following), allowing access to the GIF++ bunker. The peculiarity of the facility is the possibility of combining the muon beam and irradiation with photons, in order to study the detector response under different irradiation conditions. 

Each member of the RPC ECOgas@GIF++ collaboration provided one RPC prototype. These have been installed on two mechanical supports, located inside the GIF++ and are characterized by different layouts (area, gas gap thickness, readout system etc) as reported in table \ref{tab:detectors}.

\begin{table}[h]
\centering
\caption{Main features of the detectors from ECOgas@GIF++ collaboration}\label{tab:detectors}%
\begin{tabular}{@{}cccccc@{}}
\toprule
 & & & \textbf{Gap/electrode} & & \\
\textbf{Detector} & \textbf{Area (cm$^{\text{2}}$)}  & \textbf{Gaps} & \textbf{thickness} & \textbf{Strips} & \textbf{Readout} \\
 & & & \textbf{(mm)} & & \\
\midrule
ATLAS & 550 & 1  & 2/1.8 & 1 & Digitizer\\
LHCb/SHiP & 7000 & 1  & 1.6/1.6 & 64 & TDC\\
ALICE & 2500 & 1  & 2/2  & 32 & Digitizer\\
CMS RE11 & 1339.2+2298.5/4215.1\footnotemark[1] & 2 & 2/2 & 128 & TDC\\
BARI-1p0 & 7000 & 1 & 1/1.43 & 32 & TDC\\
EP-DT & 7000 & 1 & 2/2 & 7 & Digitizer\\
\botrule
\end{tabular}
\footnotetext[1]{The CMS RE11 RPC gas gap layout consist in 2 gaps layout labelled as Top Narrow (TN) + Top Wide (TW) and Bottom (BOT). The gap areas are therefore expressed as TN+TW/BOT}
\end{table}

The gas system allows to create the desired gas mixture (by mixing up to four gases) and sending it to the detectors. Gas flow, relative humidity and mixture composition are continuously monitored and a stop of the operations is issued, by a dedicated software, in case of wrong mixture. The high voltage is provided by means of a CAEN high voltage mainframe SY1527 \cite{mainframe}, hosting two high voltage boards (A1526N and A1526P \cite{HVboards})). 

The applied high voltage is corrected for temperature and atmospheric pressure variations, in order to maintain the effective high voltage (HV$_{\text{eff}}$) constant over time, according to the following equation, used by the CMS collaboration \cite{PTcorr}:

\begin{equation}
HV_{app} = 
HV_{eff} \left[ (1 - \alpha) + \alpha \frac{P}{P_{0}} \frac{T_{0}}{T} \right]
\label{eq:corr}
\end{equation}

where P$_{\text{0}}$ and T$_{\text{0}}$ are reference values (293.15~K and 990~mbar) and $\alpha$ is an empirical parameter set to 0.8. Note that this formula is slightly different from the one used in \cite{testBeam2021}, describing the beam tests performed in 2021. As a matter of fact, the formula used here was proved to provide better stability of efficiency and other parameters over long periods of time \cite{PTcorr}. It was not used for the 2021 data for organizational issues, since at the time not all the groups from the RPC ECOGas@GIF++ collaboration were aware and used it. Anyhow, the difference between HV$_{app}$ computed with this or the older formula is less than few tens of volts, and this makes the results presented here and the one in \cite{testBeam2021} directly comparable.

Both for aging and beam test studies, the data acquisition is carried out by the \textit{WebDCS Ecogas} \cite{webdcs}, which is a web interface to a Detector Control System, originally developed for the CMS collaboration studies at GIF++, and re-adapted to the collaboration needs. This is a versatile system, which allows the users to easily perform all the data taking, as well as produce on-the-fly data quality monitoring plots.

During beam tests, a set of scintillators (coupled with photomultipliers) is installed on the mechanical frames inside the bunker (internal scintillators) and their coincidence with two external scintillators triggers the data acquisition during the beam spill.

\begin{figure}[]
\centering
\includegraphics[width=0.75\linewidth]{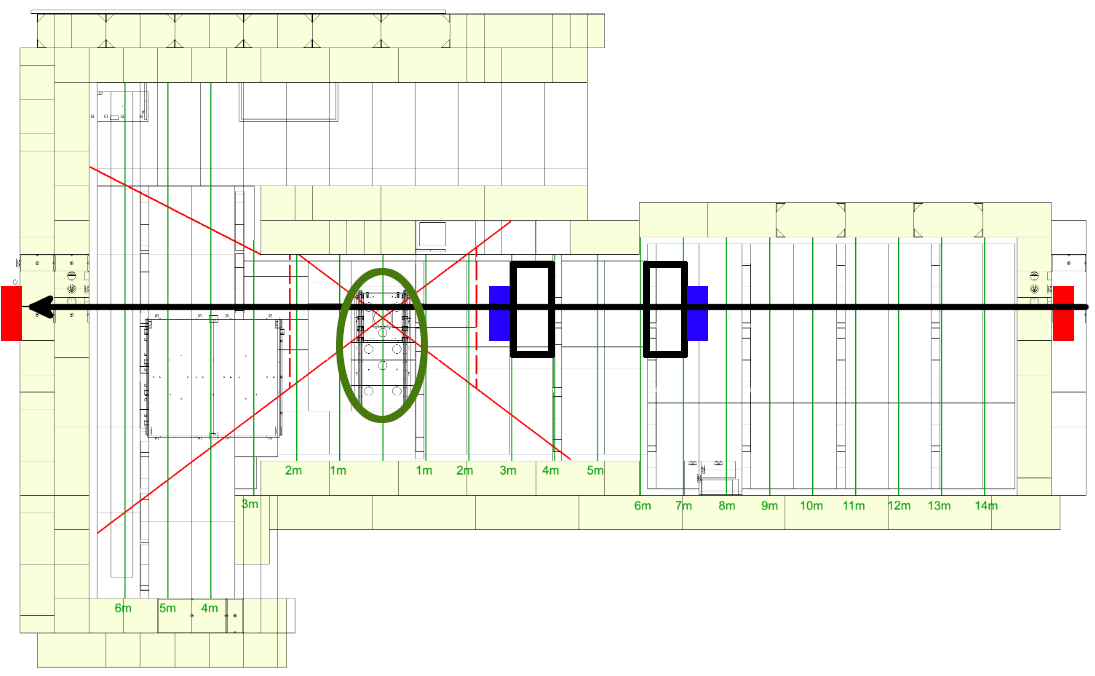}
\caption{Sketch of the GIF++ bunker with the two mechanical supports highlighted in black, internal scintillators in blue and external ones in red. The black arrow represents the direction of the muon beam. The $^{\text{137}}$Cs source is also circled in green. The red lines show the aperture of the irradiation field}\label{fig:bunker}
\end{figure}

Figure \ref{fig:bunker} shows a sketch of the GIF++ bunker, highlighting the positions of the mechanical supports (black rectangles, at $\approx$ 3 and 6 meters from the source) and of the scintillators (blue rectangles, the internal ones, and red rectangles, the external ones). The total trigger area is equal to 10$\times$10~cm$^{2}$. Lastly, the $^{137}$Cs source is also highlighted in the figure.

\section{Beam test results}
\label{sec:beamTest}

This section reports the results obtained during the 2022 beam test campaign. The first part of the section describes the detectors response when the $^{137}$Cs source is fully shielded (source-off), whilst the second part shows the behavior of the detectors when they are exposed to different values of photon fluxes (source-on).

As it is reported in table \ref{tab:detectors}, two different readout methods have been employed: for some RPCs (ALICE, EP-DT and ATLAS), the readout strips have been directly connected to a CAEN digitizer for full waveform studies (model DT5742 \cite{DT5742}\footnote{5 Gs/s 12-bit resolution} for ALICE and ATLAS, and DT5730 \cite{DT5730}\footnote{1 Gs/s and 14-bit resolution} for EP-DT) while other detectors (SHiP, CMS RE11 and BARI-1p0) use a specific front-end electronic board (FEB) to discriminate the signals which, once discriminated, are readout by means of a CAEN VME multi-hit TDC (model V1190 \cite{caenTDC}\footnote{128 channels and 100 ps time resolution on the single hit}). More details on each FEB can be found in: \cite{feeric} for the SHiP detector, \cite{ABBRESCIA2000143} for the CMS RE11, while the BARI-1p0 RPC is equipped with a custom-made 32 channels board equipped with a PETIROC ASIC \cite{PETIROC} and manufactured by Korean DEtector Laboratory (KODEL).


Mixtures with different concentrations of HFO and CO$_{2}$ were tested, while the fractions of i-C$_{4}$H$_{10}$ and SF$_{6}$ were kept constant as to study the interplay between CO$_{2}$ and HFO. The mixtures are listed in table \ref{tab:mixtures}, together with their GWP value, calculated as the average of the GWP of each component, weighted by its mass concentration, according to what is prescribed in \cite{eu-517-2014}. The GWP can be used to compare the effects of the same mass of gas, expelled into the atmosphere. However, the RPC gas systems at the LHC operate at a fixed rate of gas volume changes; for this reason, to compare different mixtures (with different specific masses), one can introduce the CO$_{2}$-equivalent (CO$_{2}$e) for 1~liter of gas mixture released into the atmosphere, expressed in grams per liter. These values are reported in the last column of table \ref{tab:mixtures}. Note that the first mixture in the table (STD) does not contain any HFO (or CO$_{2}$): it represents the standard gas mixture, currently employed in the ATLAS/CMS RPCs and it has been taken as a reference to which the eco-friendly alternatives are compared.

\begin{table}[h]
\caption{Composition of the gas mixtures used in the tests described in this paper}\label{tab:mixtures}%
\begin{tabular}{@{}cccccccc@{}}
\toprule
\textbf{Mixture} & \textbf{C$_{2}$H$_{2}$F$_{4}$ \%} & \textbf{HFO \%} & \textbf{CO$_{2}$ \%} & \textbf{i-C$_{4}$H$_{10}$ \%} & \textbf{SF$_{6}$ \%} & \textbf{GWP} & \textbf{CO$_{2}$e (g/l)} \\
\midrule
            \textbf{STD} & 95.2 & 0 & 0 & 4.5 & 0.3 & 1485 & 6824\\
			\textbf{MIX0} & 0 & 0 & 95 & 4 & 1 & 730 & 1480\\
			\textbf{MIX1} & 0 & 10 & 85 & 4 & 1 & 640 & 1490\\
			\textbf{MIX2} & 0 & 20 & 75 & 4 & 1 & 560 & 1495 \\
			\textbf{MIX3 or ECO3} & 0 & 25 & 69 & 5 & 1 & 527 & 1519 \\
			\textbf{MIX4} & 0 & 30 & 65 & 4 & 1 & 503 & 1497\\
			\textbf{MIX5 or ECO2} & 0 & 35 & 60 & 4 & 1 & 476 & 1522\\
			\textbf{MIX6} & 0 & 40 & 55 & 4 & 1 & 457 & 1500\\
\botrule
\end{tabular}
\end{table}

It is worth noting that the CO$_{2}$e for all the mixtures tested is quite similar; this is due to the fact that the SF$_{6}$ concentration is the same and it is the only GHG in the mixtures tested and that they are $\approx$4/4.5 times lower with respect to the standard gas mixture one.

\subsection{Results without gamma background}
\label{subsub:sourceOff}

This section summarizes the main results that have been obtained during the 2022 test beam campaigns, carried out by the RPC ECOgas@GIF++ collaboration. In particular, the results obtained when the $^{137}$Cs source is fully shielded and no background radiation is present on the detectors (source-off) are reported here. 

\subsubsection{Source off Efficiency and Working Point}
\label{subsub.eff}

\begin{figure}[]
\centering
\includegraphics[width=\linewidth]{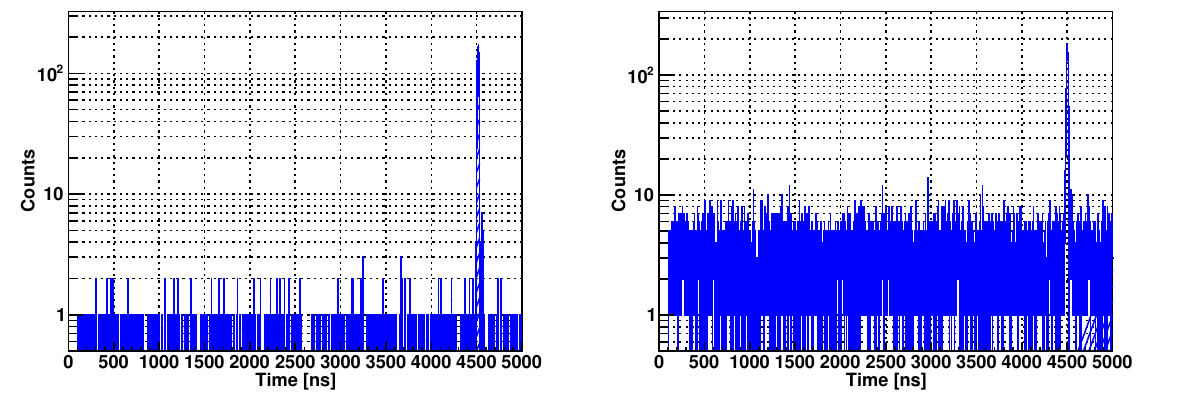}
\caption{Hit time profile obtained with the CMS RE11 RPC when flushed with the STD gas mixture at 90\% efficiency. Left panel: source-off condition. Right panel: highest possible (irradiator fully opened) $\gamma$ background condition}
\label{fig:timeProfile}
\end{figure}

Both the digitizers and the TDCs provide a timestamp for each hit they register, regardless of its origin. This time information can be used to separate the muon-induced hits from those coming from other sources, such as noise and/or the $\gamma$ background. Figure \ref{fig:timeProfile} shows the time profiles (distribution of the hits arrival times for a fixed high voltage value) obtained for the RE11 RPC, operated with the standard gas mixture at 90\% efficiency at source-off (left panel) and under the highest possible (irradiator fully opened) $\gamma$ background condition (right panel).

In the case of the TDCs, the width of the acquisition window was set to 5000~ns while in the case of the digitizer to: 500~ns for the ATLAS detector, and 1040~ns and 1024~ns for the EP-DT and ALICE chambers, respectively. The different widths of each acquisition window have been taken into account while performing the analysis (as it will be explained later on in section \ref{subsub:chargeDistr}). The choice of TDC or digitizer was made due to different requirements of the groups involved (i.e. the digitizer is used to study the analog response of the detector while the TDCs, coupled with front-end electronics, are used to simulate real-life conditions for RPCs in the LHC). The peaks that are clearly visible in both panels (with a width of $\approx$25~ns) of figure \ref{fig:timeProfile} correspond to an accumulation of muon-induced events (since the time interval between the trigger and the muon hit is the same for each event) while the others, uniformly distributed, are due to the noise/gamma-induced hits. 

\begin{figure} []
    \centering
    \subfloat[\empty]{
        {\includegraphics[width=0.49\linewidth]{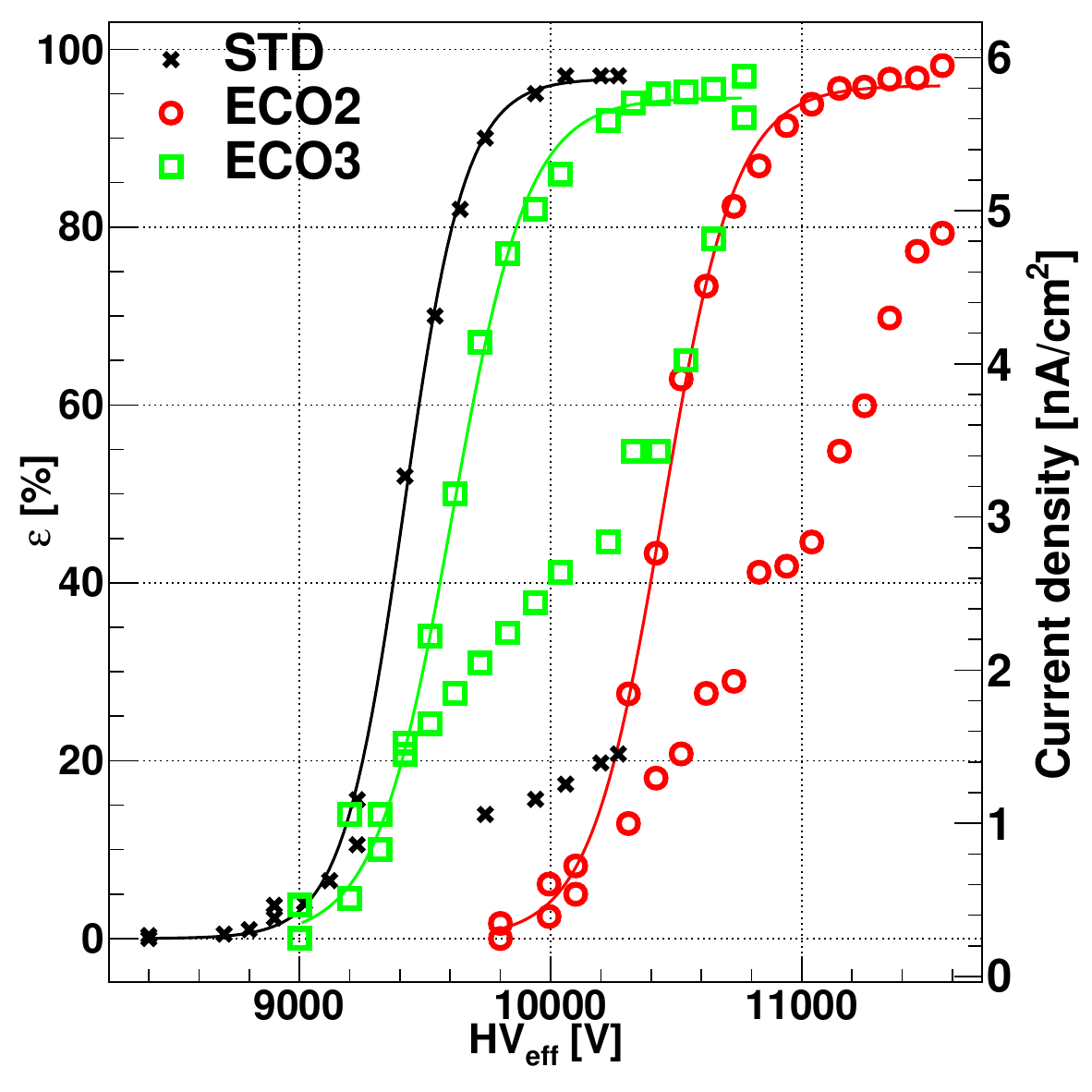}}
        \label{fig:effATLASsmall}
        }
    \subfloat[\empty]{
        {\includegraphics[width=0.49\linewidth]{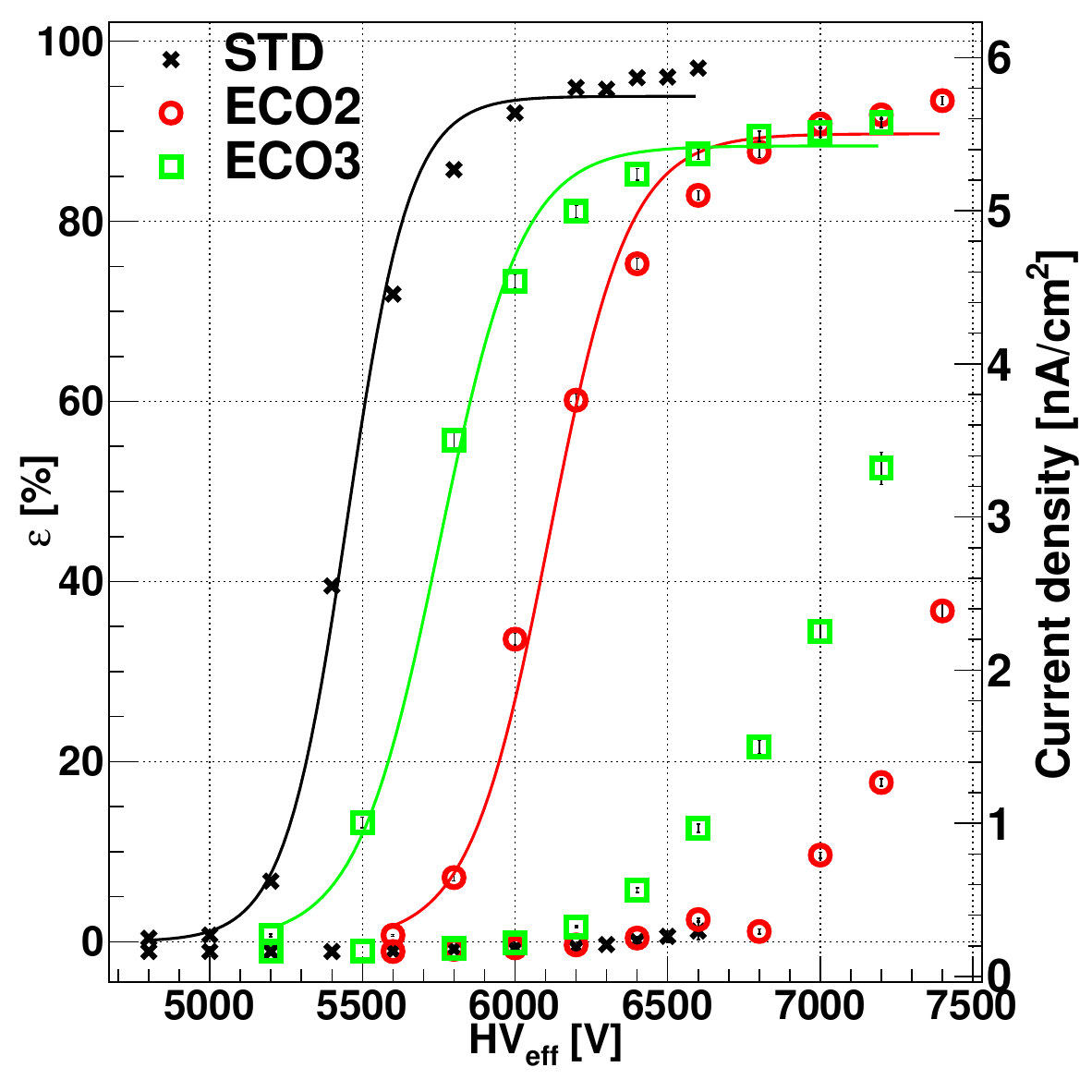}}
        \label{fig:effBARI1p0}
        }
    \caption{Efficiency and absorbed current density as a function of HV$_{eff}$, without $\gamma$ background. Left panel: ATLAS RPC. Right panel: BARI-1p0 RPC}
    \label{fig:effATLASBARI}
\end{figure}

The chamber efficiency can then be computed as the ratio between the number of events whose time falls inside the muon window and the number of triggers. The left and right panels in figure \ref{fig:effATLASBARI} show the efficiency and absorbed current density (reported in~nA/cm$^{2}$, to compare it across detectors with different active areas) as a function of the effective high voltage applied to the detectors without $\gamma$ background for three different mixtures: STD, ECO2 (35/60~HFO/CO$_{2}$) and ECO3 (25/69~HFO/CO$_{2}$) for the ATLAS (2~mm single gas gap) and the BARI-1p0 (1~mm single gas gap) respectively.

The efficiency data points were interpolated using the logistic function reported in equation \ref{eq:fitEff}:

\begin{equation}
\varepsilon(HV) = \frac{\varepsilon_{max}}{1+e^{-\beta(HV-HV_{50})}}
\label{eq:fitEff}
\end{equation}


where the free parameters are: $\varepsilon_{max}$, which represents the asymptotic maximum efficiency (plateau efficiency); $\beta$, which is related to the steepness of the efficiency curve, and HV$_{50}$, which represents the voltage where the efficiency reaches 50\% of its maximum. These values are used to compute the voltage corresponding to the Working Point (WP), according to the definition used by the CMS collaboration \cite{cmsRPC}:

\begin{equation}
\text{WP} = \frac{\log{19}}{\beta} + HV_{50} + \text{150 V}
\label{eq:WP}
\end{equation}

Table \ref{tab:parametersEffOff} reports the most significant parameters for the detectors shown in figure \ref{fig:effATLASBARI}. It is possible to observe that, in the case of the ATLAS RPC, the plateau efficiency is compatible for the different mixtures while, in the case of a thinner gap, such as the BARI-1p0 detector, the efficiency decreases when the eco-friendly alternatives are used. A possible explanation could be that the CO$_{2}$ produces a smaller number of primary ion-electron clusters due its lower density \cite{sauli}, leading to a more significant efficiency drop in thinner gas gaps.
Furthermore, as anticipated in section \ref{sec:intro}, the inclusion of HFO to the mixture tends to move the detector WP to higher values. The increase with respect to the standard gas mixture is similar for both detectors for ECO3 (around 0.3/0.4~kV) while for ECO2 it is around 1~kV for the 2 mm gas gap detector and around 0.8~kV for the 1~mm gas gap RPC.

\begin{table}[h]
\caption{Source off WP and plateau efficiency for the ATLAS and the BARI-1p0 RPCs}\label{tab:parametersEffOff}%
\begin{tabular}{@{}cccc@{}}
\toprule
\textbf{Mixture} & \textbf{Detector} & \textbf{WP [V]} & \textbf{$\varepsilon_{max}$ [\%]}  \\
\midrule
        \textbf{STD} & ATLAS & 9925.7 & 96.71 \\
		\textbf{ECO2} & ATLAS & 11021.9 & 95.92 \\
		\textbf{ECO3} & ATLAS & 10200.7 & 94.56 \\
        \textbf{STD} & BARI-1p0 & 5903 & 93.89 \\
		\textbf{ECO2} & BARI-1p0 & 6646.2 & 89.73 \\
		\textbf{ECO3} & BARI-1p0 & 6301.2 & 88.38 \\
\botrule
\end{tabular}
\end{table}

\subsubsection{Signal charge distribution and large signals contamination}
\label{subsub:chargeDistr}
By using a digitizer, one has access to the waveform of each signal detected by the RPCs under test and this allows to perform a full characterization of the detector response, especially in terms of signal charge and contamination from large signals. The starting point in the signal charge calculation is the discrimination between signals and noise. In the following, results from EP-DT and ALICE detectors are presented, hence a few details on the procedure employed in the analysis are described here. In the case of the EP-DT detector, a waveform is considered to contain a signal from a muon if its amplitude is above 2~mV with respect to the baseline while in the case of ALICE, the threshold was set to five times the RMS of the signal in a region where no muon signal is expected (noise window). 

\begin{figure}[]
\centering
\includegraphics[width=0.4\linewidth]{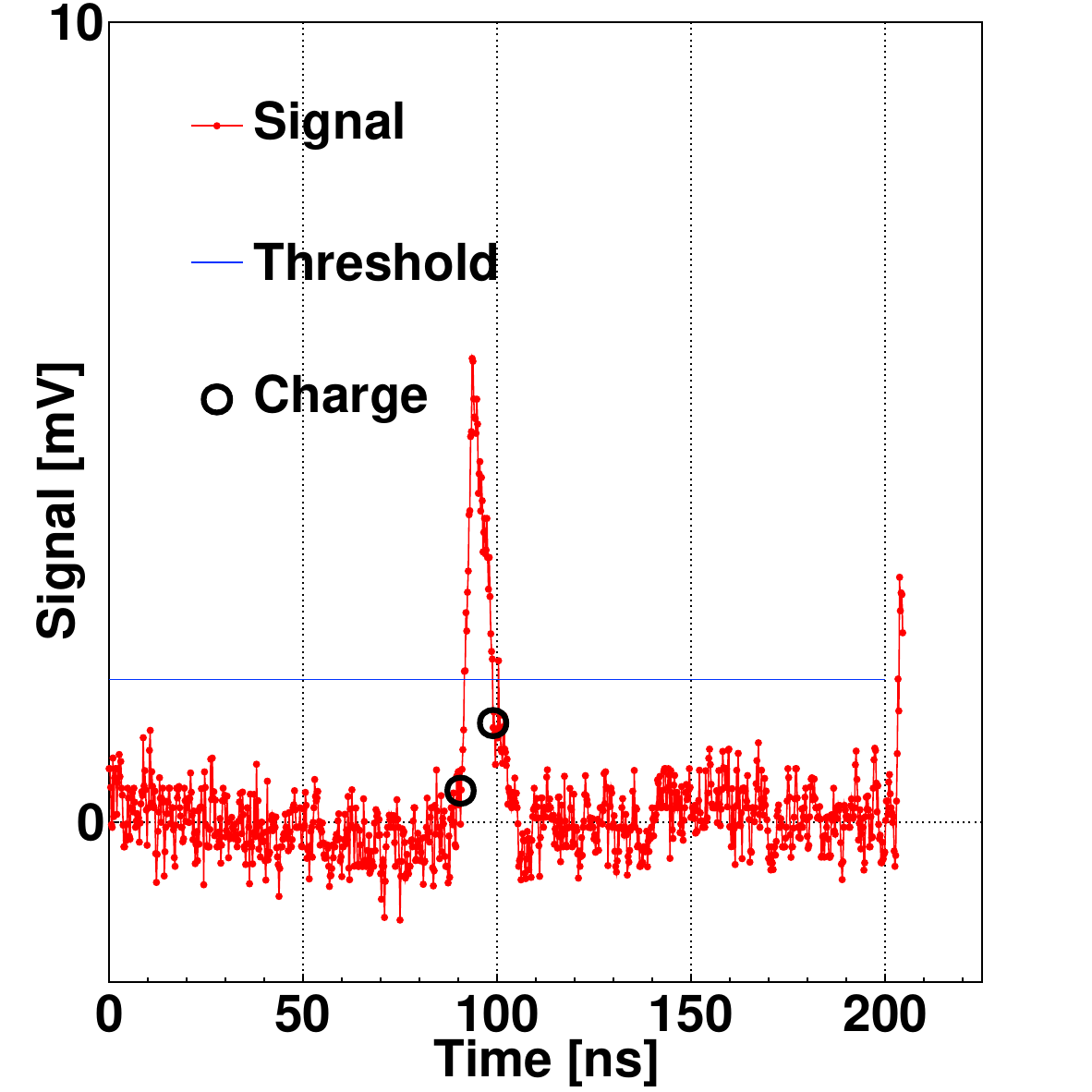}
\caption{Example of muon signal recorded by the digitizer of the ALICE RPC. The black circles refer to the start/end of the integration interval for signal charge calculation and the blue line represent the threshold}\label{fig:exSignal}
\end{figure}

The signal charge is then calculated by integrating the signals passing the above selection criterion in a suitable integration window. In particular, the range for signal integration is determined as follows (a visual reference is also reported in figure \ref{fig:exSignal}):

\begin{itemize}
    \item The first and last samples where the signal is above threshold are determined for each strip

    \item Starting from the first (last) point the signal is swept forward (back) and the discrete derivative between two consecutive samples is calculated. When this changes sign, it means that there is a change in the signal slope and the last point before the sign change is assumed to be the start (end) point of the integration interval (for a complete description of the algorithm the reader can refer to \cite{PhDGianluca, PhDLuca})

    \item The charge calculated for each strip is then summed, to get the total charge per event
\end{itemize}

Figure \ref{fig:exSignal} shows an example of a signal as seen by the ALICE RPC, when flushed with the standard gas mixture: the horizontal blue line represents the threshold in the specific event and the two black markers show the start and end of the integration interval just described. Since the signal is readout on a 50~$\Omega$ resistor, to find the value of signal charge, the result of the signal integration is divided by 50~$\Omega$.

\begin{figure} []
    \centering
    \subfloat[\empty]{
        {\includegraphics[width=0.49\linewidth]{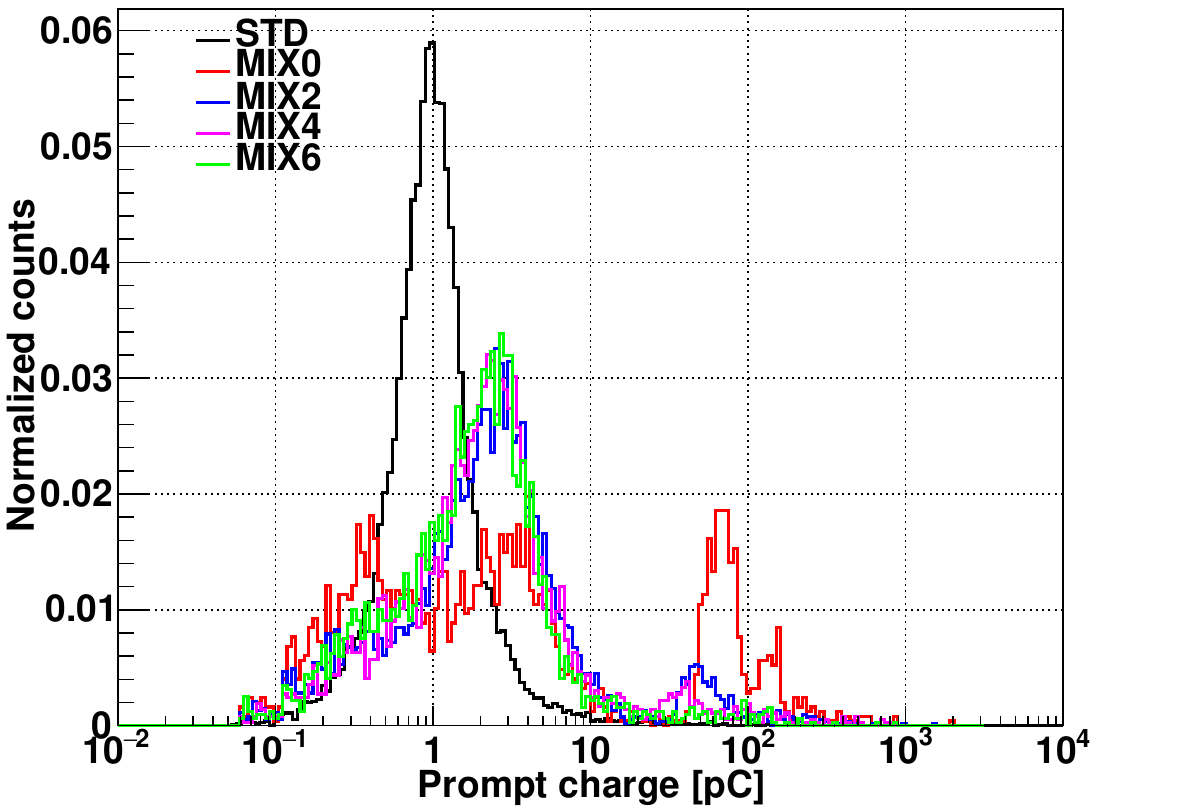}}
        \label{fig:chargeEPDT}
        }
    \subfloat[\empty]{
        {\includegraphics[width=0.49\linewidth]{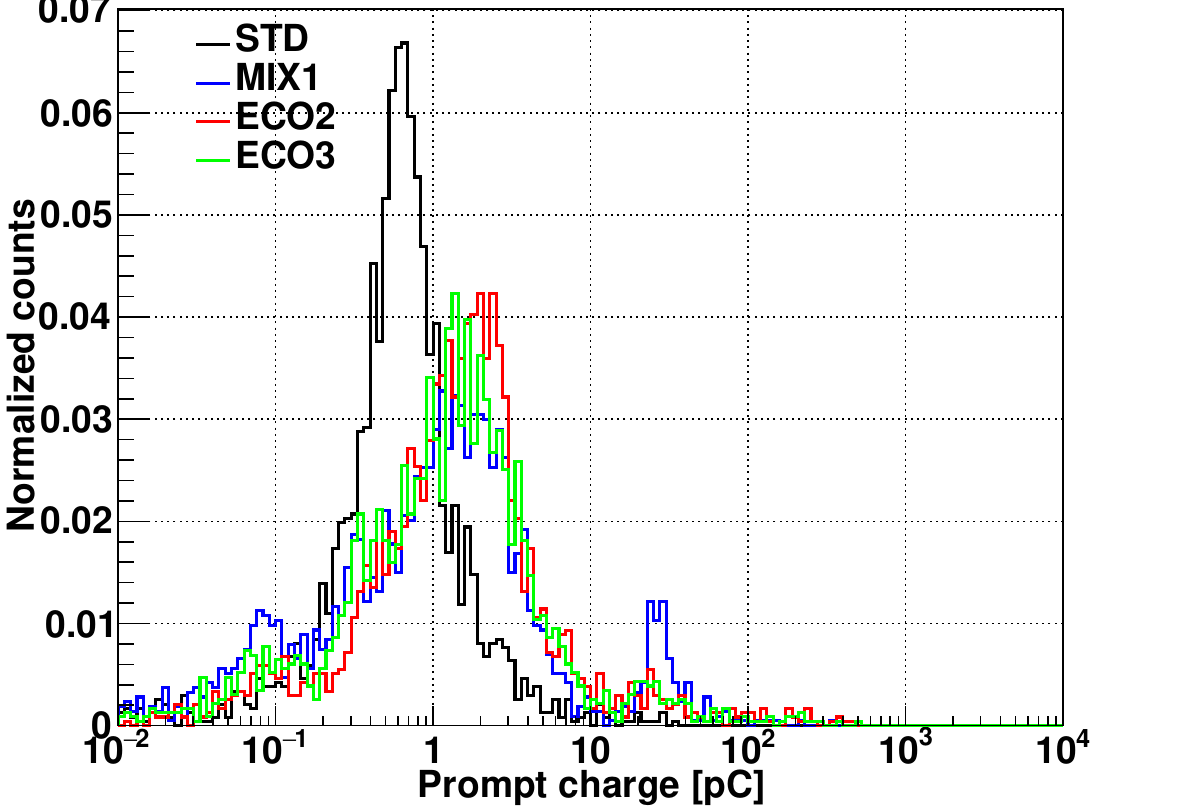}}
        \label{fig:chargeALICE}
        }
    \caption{Signal charge distributions at the calculated working point without $\gamma$ irradiation. Left panel: EP-DT RPC. Right panel: ALICE RPC. Note that different gas mixtures are used in the two detectors}
    \label{fig:chargeWP}
\end{figure}

The left and right panels of figure \ref{fig:chargeWP} show, respectively, the charge distributions for the EP-DT and ALICE detector when the applied high voltage is the closest to the estimated working point for the tested gas mixtures. Note that the average threshold used for the ALICE detectors ($\approx$1.6~mV) is lower with respect to the EP-DT one (2~mV), and this likely affects the average values of the signal charges measured in the two cases. Indeed, it appears that for the ALICE detector, the average charge values are slightly lower, with respect to the EP-DT one.

In both cases, the black distribution in figure \ref{fig:chargeWP} refers to the standard gas mixture, while the others refer to all the tested eco-friendly alternatives. For all mixtures, two peaks can be observed, the one at lower charge values coming from the avalanche contribution while the one at higher charge values from larger signals. Usually, these are referred to as streamers \cite{avStreamer} but, in the case of the eco-friendly alternatives, not all signals in the right peak of the distribution in figure \ref{fig:chargeWP} satisfy all the criteria to be defined streamers (i.e. they are not always accompanied by a precursor signal and they might be characterized by multiple delayed peaks). For this reason, these will be referred to as "large signals" from here on. In general, for the eco-friendly alternatives, the avalanche peak is shifted towards higher values with respect to the standard gas mixture (this observation is consistent with the higher absorbed current, as reported in \ref{subsub.eff} and figures therein). Moreover, the fraction of large signals is generally larger, although this value seems to be lower by increasing HFO concentration in the mixtures. But, the decrease of these kinds of signals comes with the price of a higher working point, mainly due to the quenching effect of adding more HFO to the mixture.

\begin{figure} [] 
    \centering
    \subfloat[\empty]{
        {\includegraphics[width=0.49\linewidth]{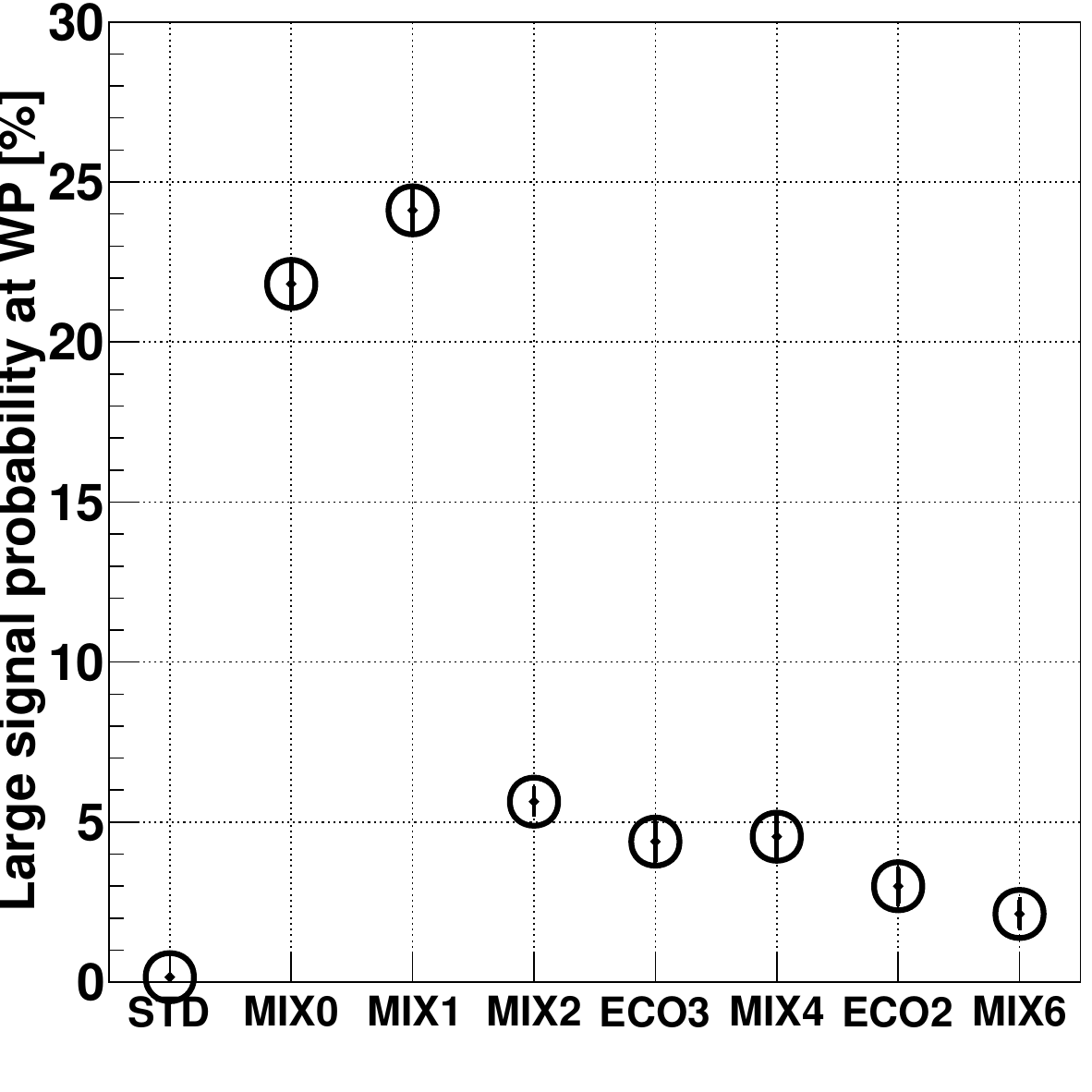}}
        \label{fig:strProbEPDT}
        }
    \subfloat[\empty]{
        {\includegraphics[width=0.49\linewidth]{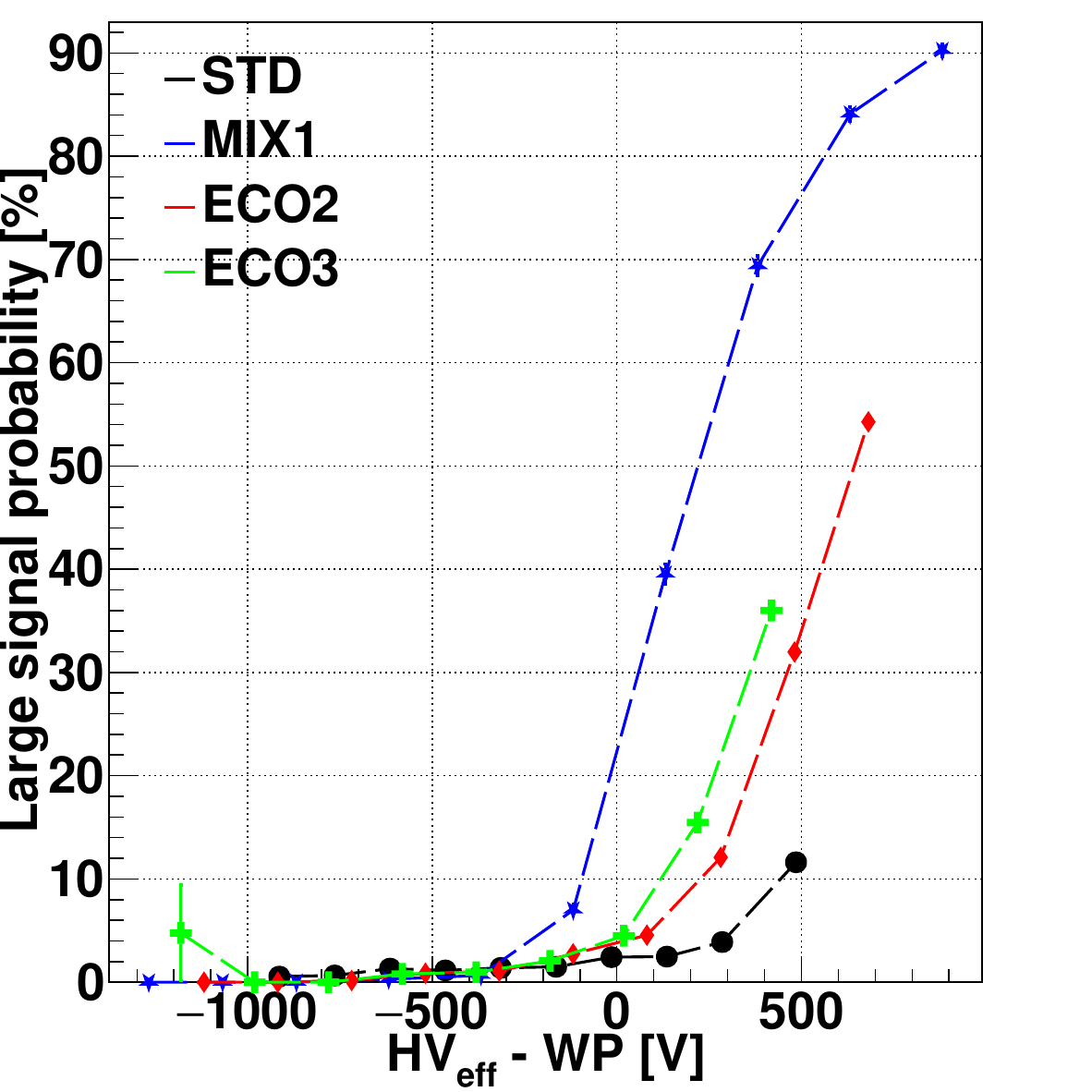}}
        \label{fig:strProbALICE}
        }
    \caption{Left panel: Large signal probability at WP for the EP-DT detector. Right panel: large signal probability as a function of HV$_{eff}$ for the ALICE detector}
    \label{fig:streamers}
\end{figure}

In order to quantify this value, all events characterized by a charge larger than 16~pC are tagged as "large signals" (this value was chosen by observing that the two peaks in the charge distributions are separated at $\approx$ the 16~pC mark). The large signal probability can then be defined as the ratio between the number of these signals and the total number of events. Figure \ref{fig:streamers} shows, in the left panel, the values of large-signal probability at the working point, for the EP-DT detector, while the right panel shows the large-signal probability as a function of the applied high voltage, in the case of the ALICE detector. It is possible to observe how, for increasing HFO concentration, the contamination from large signals at WP reaches similar values as the standard gas mixture but it tends to increase more sharply for voltages above the WP. This leads to a narrower range of applicable high voltage which grants both a high detection efficiency ($>$ 95\%) as well as a low large signal probability ($<$ 5\%). The large signals contamination for the ALICE detector is reported as a function of the high voltage minus working point; in this way, the point at 0~V corresponds to the WP for all mixtures.

\subsubsection{Source off summary}
\label{subsub:sourceOffResults}
Using the previously shown results, a few conclusions can be drawn. First of all, for thinner gas gap detectors (such ad the BARI-1p0), an HFO fraction above 50\% is advisable, in order to reach a high enough efficiency plateau. This is less true for the 2~mm detectors (as in the case of the ATLAS RPC), where mixtures with 25\% HFO already provide an efficiency $>$95\% at working point. The detector working point also increases if the HFO concentration increases, at a level of $\approx$1~kV for every 10\% HFO added to the mixture. Lastly, by studying the signal charge and large signals contamination, it is worth noting that the average signal charge is higher for all the eco-friendly alternatives, leading to the higher absorbed currents observed. Moreover, for increasing HFO fractions, the average avalanche charge and large signal contamination both decrease. 

\subsection{Results with gamma background}
\label{subsub:sourceOn}

\begin{figure} [!h] 
    \centering
    \subfloat[\empty]{
        {\includegraphics[width=0.33\linewidth]{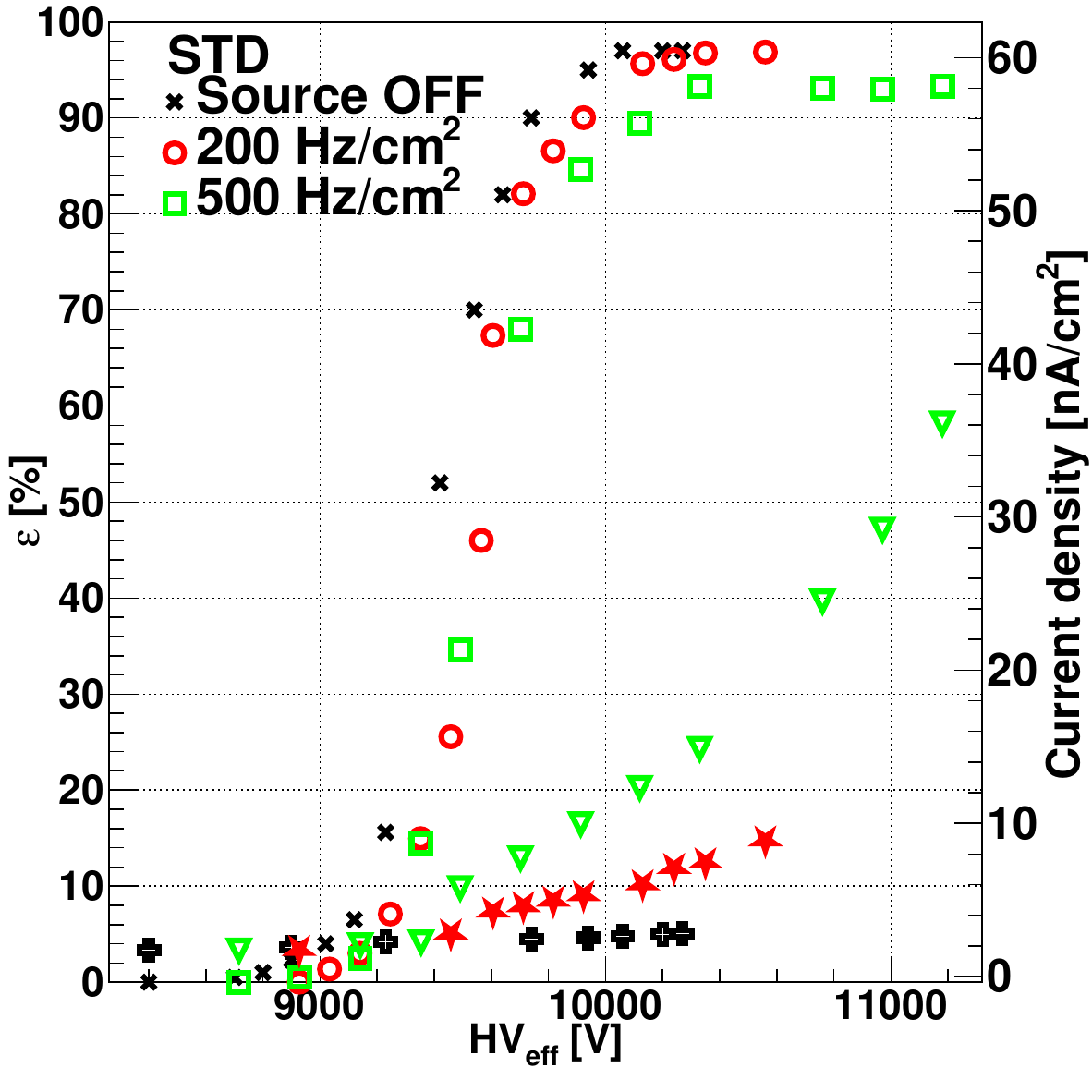}}
        \label{fig:atlasOnHVSTD}
        }
    \subfloat[\empty]{
        {\includegraphics[width=0.33\linewidth]{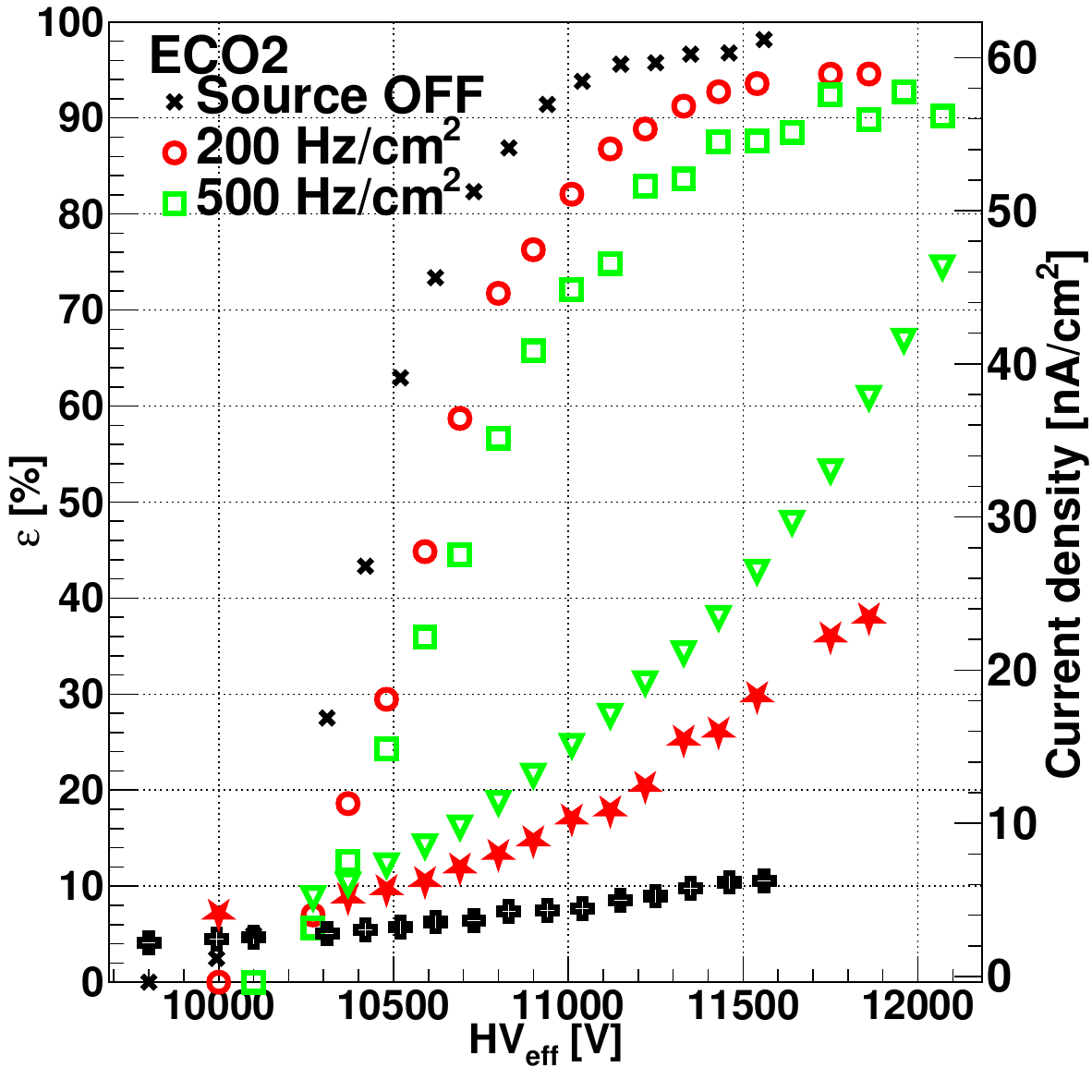}}
        \label{fig:atlasOnHVECO2}
        }
    \subfloat[\empty]{
        {\includegraphics[width=0.33\linewidth]{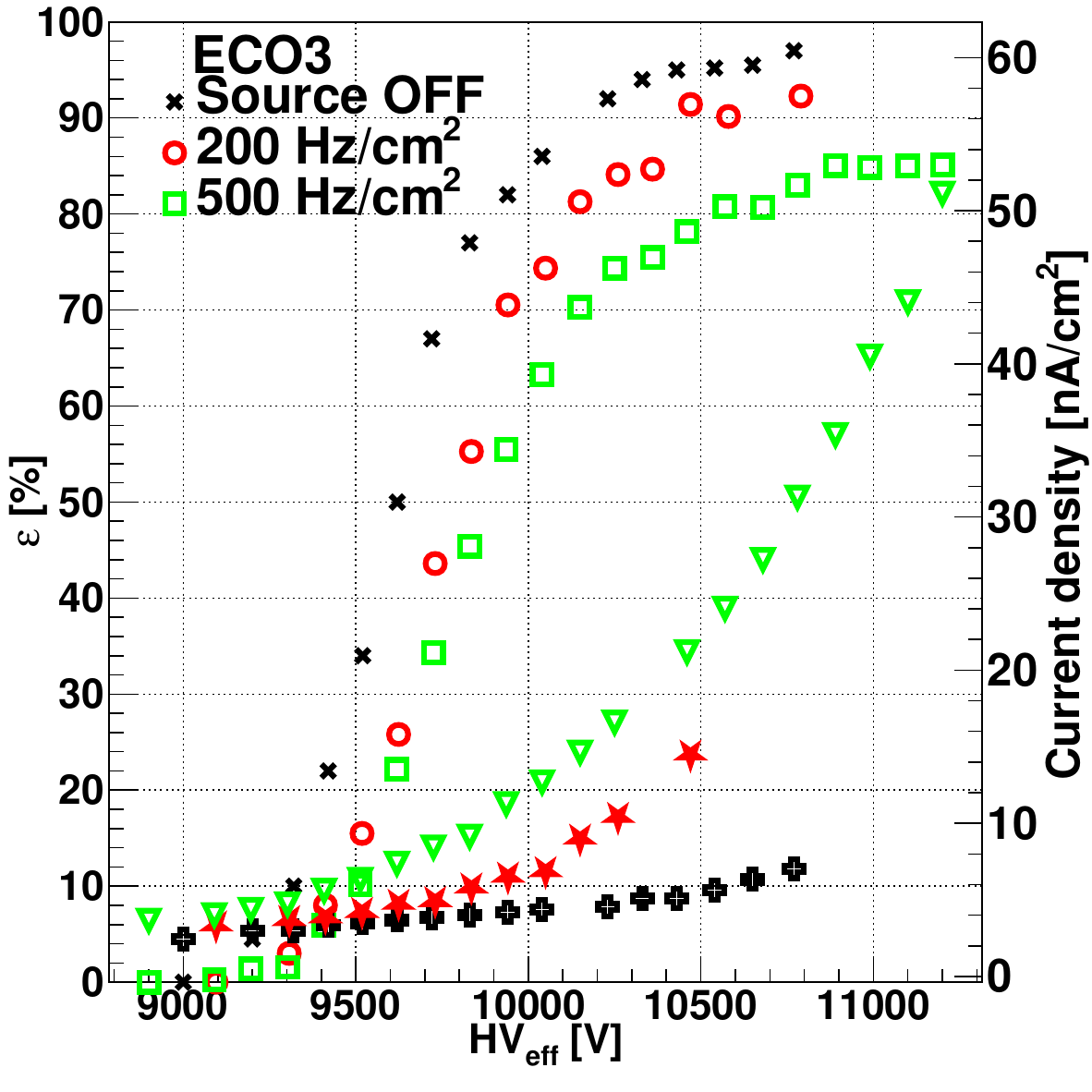}}
        \label{fig:atlasOnHVECO3}
        }
    \caption{Efficiency and current densities for the ATLAS detector in different regimes of irradiation. Left panel: STD mixture. Central panel: ECO2 mixture. Right panel: ECO3 mixture}
    \label{fig:atlasOnHV}
\end{figure}

The performance of the chambers under test were also investigated using different combinations of attenuation filters to shield the $^{137}$Cs source. Figure \ref{fig:atlasOnHV} shows the efficiency and absorbed current density, as a function of HV$_{eff}$, for the ATLAS detector (located at $\approx$3~m from the source) with the STD gas mixture (left panel) and two HFO-based candidates, ECO2 (middle panel) and ECO3 (right panel), in different conditions of $\gamma$ background.

The efficiency curves with the three mixtures are shifted at higher voltages for increasing $\gamma$ rates. This phenomenon occurs because of the increased $\gamma$ background, which leads to a higher absorbed current. Flowing through the resistive chamber electrodes, this current leads to a voltage drop across them which, in turns, leads to a reduction of the electric field inside the gap, which must be recovered by increasing the supplied high voltage. Moreover, another effect, which will be better described in \ref{subsub:effSourceOn}, can also be observed: the $\epsilon_{max}$ reached decreases if the irradiation level is increased. The right panel of figure \ref{fig:atlasOnHV} shows that for the ECO3 gas mixture, the efficiency decrease due to the $\gamma$ background is the highest among the three. For what concerns the current densities, those follow the expected behavior, increasing with the $\gamma$ background rate. The highest increase is observed if the CO$_2$ concentration increases (as it is the case for the ECO3 mixture).

\subsubsection{Gamma cluster rate}

\begin{figure} [!h] 
    \centering
    \subfloat[\empty]{
        {\includegraphics[width=0.49\linewidth]{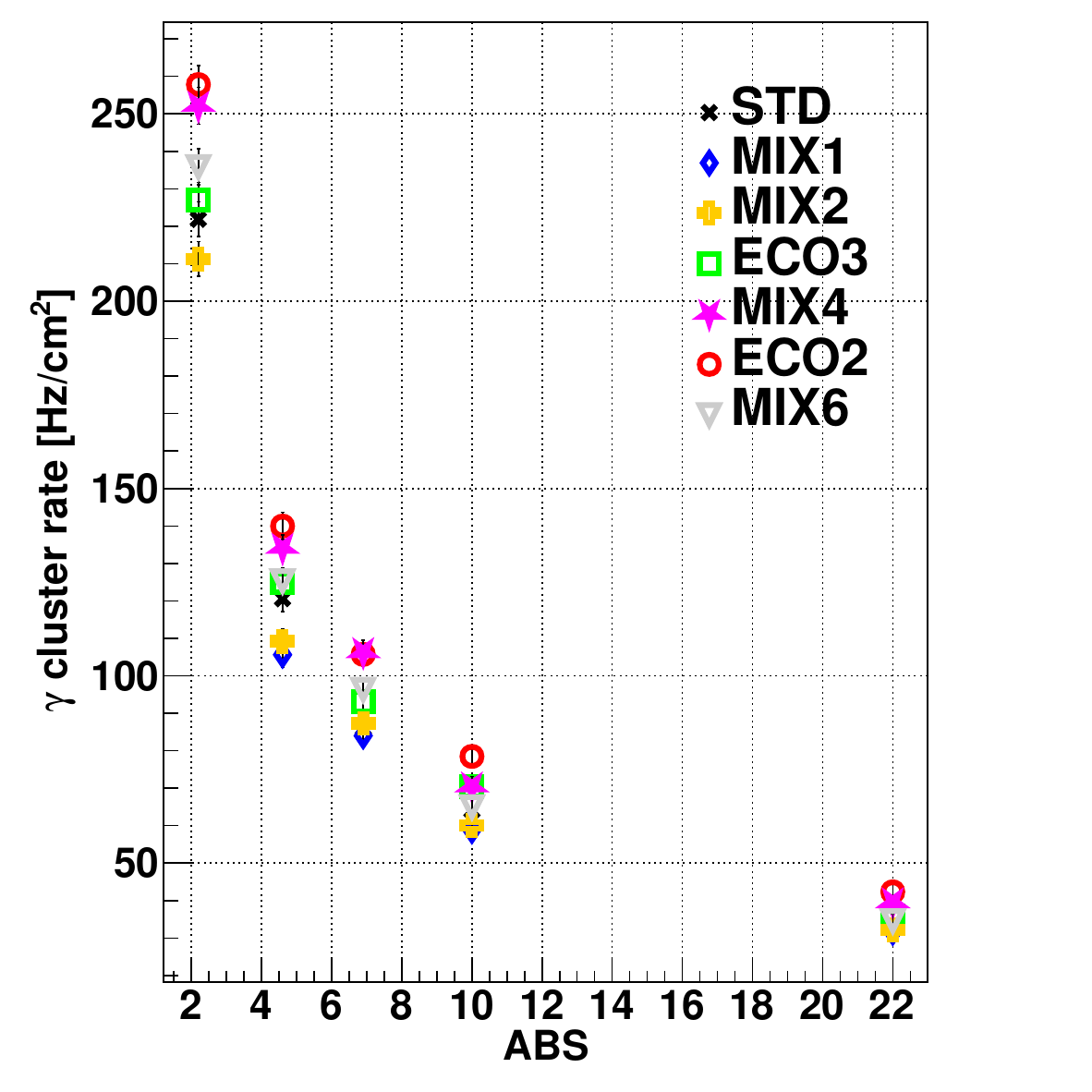}}
        \label{fig:AliceRateAbs}
        }
    \subfloat[\empty]{
        {\includegraphics[width=0.49\linewidth]{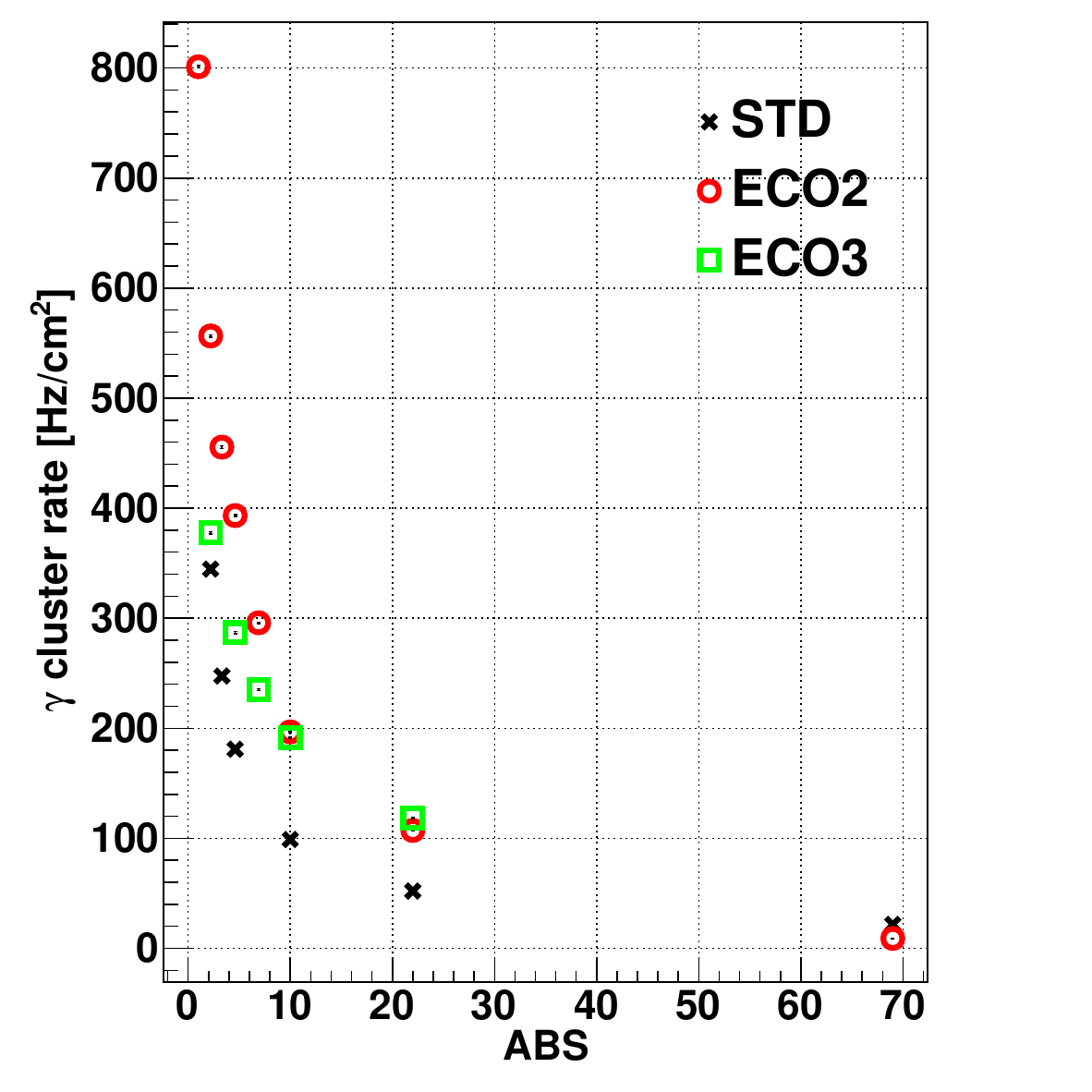}}
        \label{fig:bariRateAbs}
        }
    \caption{Gamma cluster rate at working point as a function of the attenuation filter configuration (ABS). Left panel: ALICE RPC. Right panel: BARI-1p0 RPC}
    \label{fig:rates2mm}
\end{figure}

The $\gamma$ cluster rate was measured for all the detectors with different absorption factors. This approach allows to assess the rate capabilities of each RPC at specific distances from the source. The $\gamma$ cluster rate is calculated using the data collected when no beam is present. The RPC response is sampled using a random trigger (a pulse sent to the DAQ modules with a given frequency) and, for each trigger, the data is grouped in clusters (i.e. a $\gamma$ can lead to an above-threshold signal on more than one adjacent strip). The number of $\gamma$ clusters is then counted and divided by the total acquisition time (5000 ns $\times$ number of random triggers) multiplied by the detector active area (to get a measure in Hz/cm$^{2}$). It has to be noted that the absolute photon rate varies as $1/r^{2}$ (r is the distance from the source) but one needs to consider that the actual $\gamma$ rate measured by the RPCs is affected by other factors, such as the absorption of photons from other setups in front of a given RPC and the intrinsic $\gamma$ detection efficiency of each detector. For these reasons, one cannot expect to find the $1/r^{2}$ dependence only by looking at the measured $\gamma$ rate. Figure \ref{fig:rates2mm} shows the $\gamma$ cluster rates measured at the WP for the ALICE and BARI-1p0 detectors (located at 6 and 3~m from the source respectively). 
The values are in agreement with the different distances from the source and gap features of the two RPCs. 




\subsubsection{Efficiency and Working Point}
\label{subsub:effSourceOn}
\begin{figure} [!h] 
    \centering
    \subfloat[\empty]{
        {\includegraphics[width=0.49\linewidth]{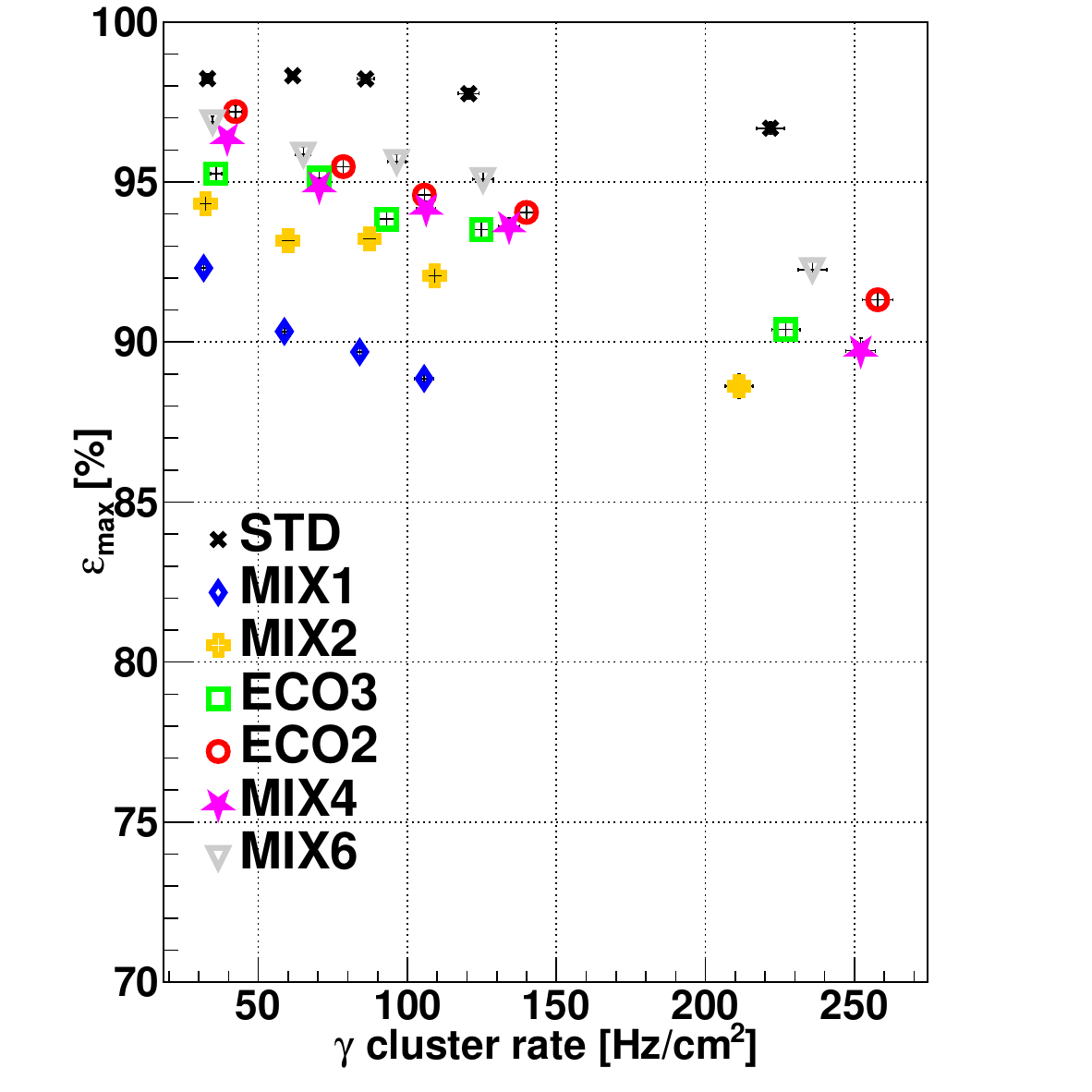}}
        \label{fig:AliceEffRate}
        }
    \subfloat[\empty]{
        {\includegraphics[width=0.49\linewidth]{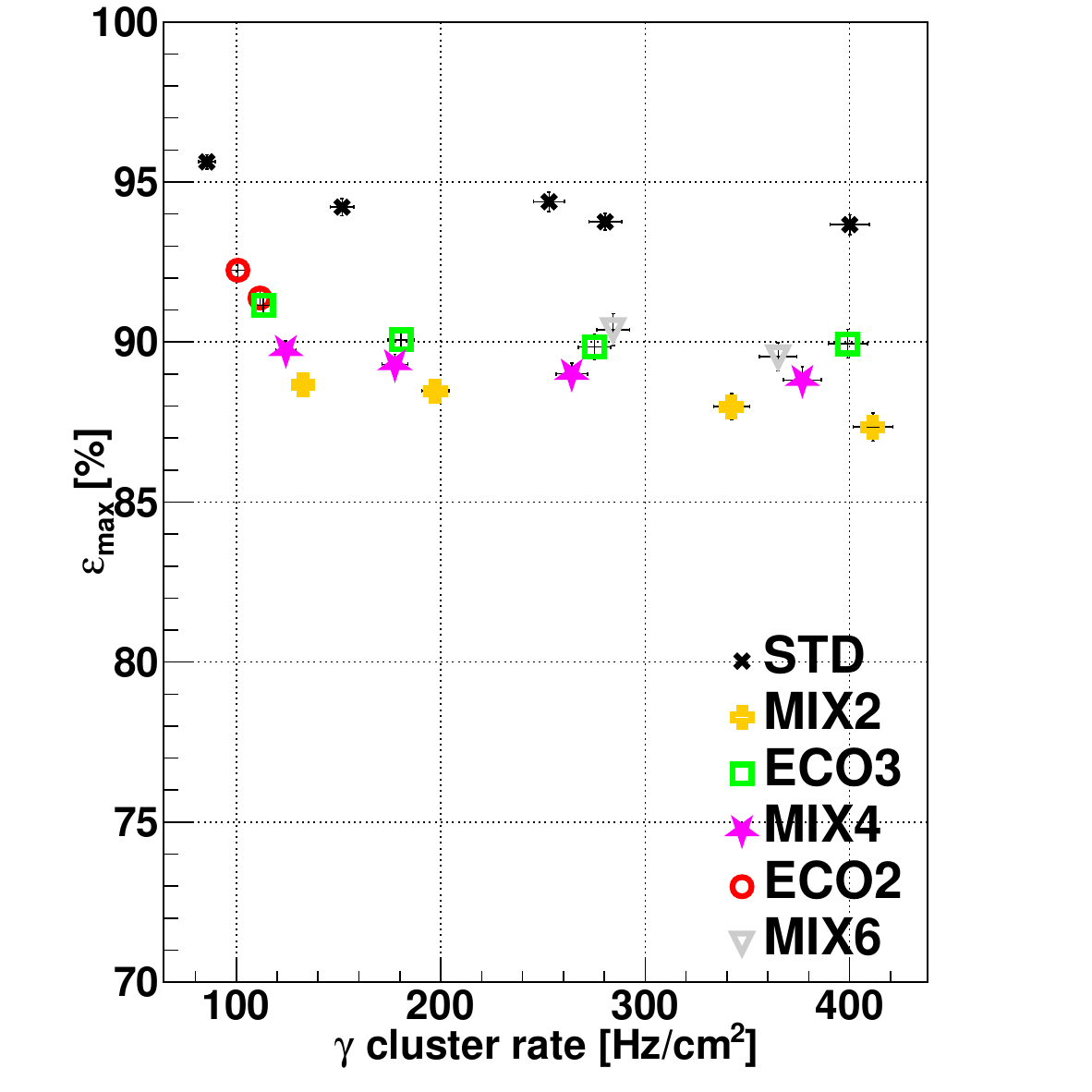}}
        \label{fig:epDtEffRate}
        }
    \caption{Plateau efficiency ($\epsilon_{max}$) as a function of the $\gamma$ cluster rate measured at WP. Left panel: ALICE detector. Right panel: EP-DT detector}
    \label{fig:eff2mm}
\end{figure}

The efficiency under irradiation was measured following the method described in section \ref{subsub.eff}. Figure \ref{fig:eff2mm} shows the values of maximum efficiency ($\epsilon_{max}$) as a function of the measured $\gamma$ cluster rate for the ALICE and EP-DT detectors operated with different mixtures. 

The same trend already shown for the ATLAS detector in figure \ref{fig:atlasOnHV} is also visible in this case: $\epsilon_{max}$ decreases if the irradiation increases. This can be explained by considering what was already anticipated earlier regarding the reduction of the high voltage effectively applied to the gas when the RPCs are exposed to a high $\gamma$ flux, due to the voltage drop across the resistive electrodes of the detectors. The highest efficiencies (and the smallest decrease for increasing background) for both detectors are obtained using the STD mixtures, while with the eco-friendly candidates the efficiency decrease at higher rates is more pronounced with respect to the STD mixture; a similar behavior is observed for the ALICE and EP-DT detectors.

\begin{figure}[!h]
\centering
\includegraphics[width=0.49\linewidth]{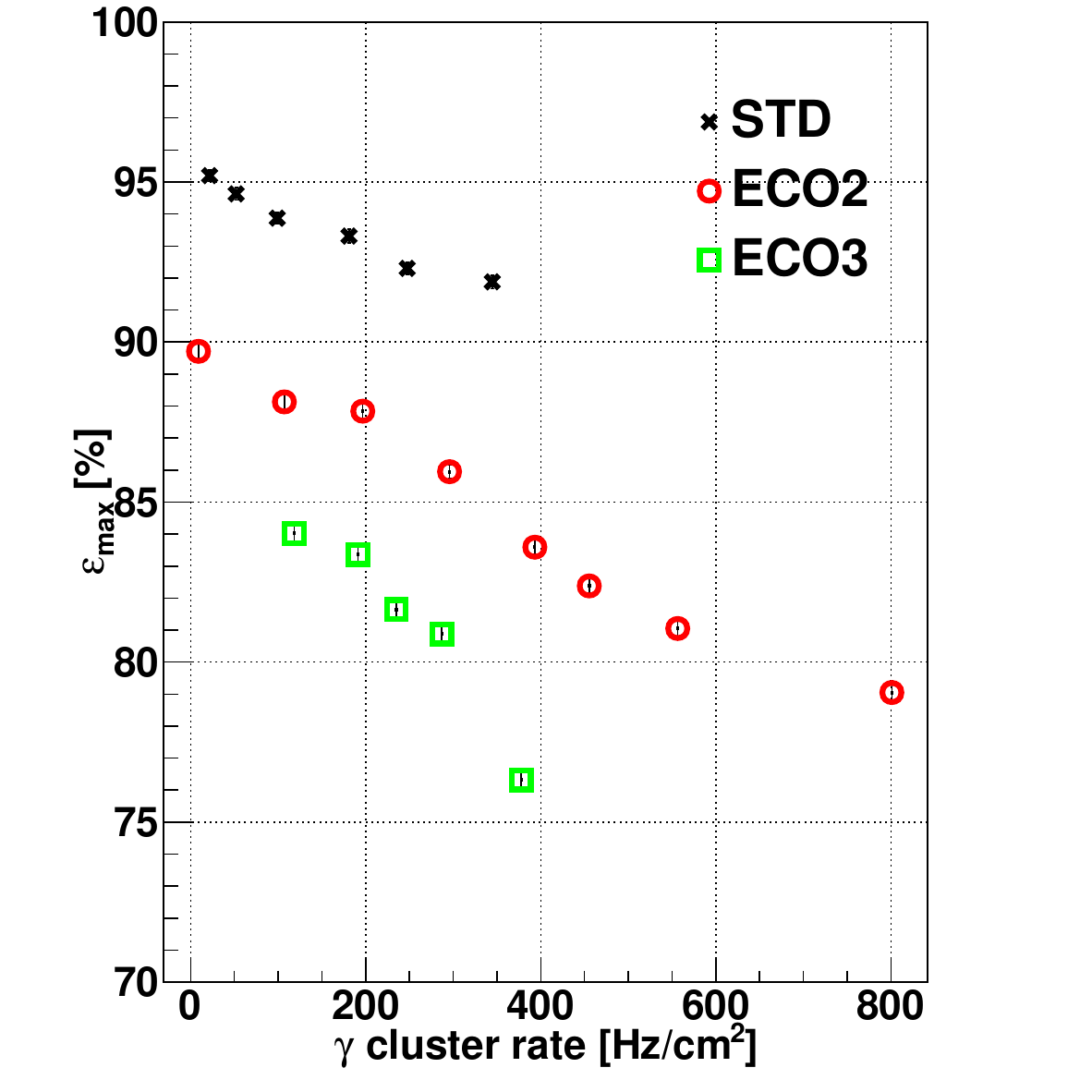}
\caption{Plateau efficiency ($\epsilon_{max}$) for the BARI-1p0 RPC as a function of the $\gamma$ cluster rate measured at WP}\label{fig:bariEffRate}
\end{figure}

The plateau efficiency measured with the BARI-1p0 detector at different $\gamma$ cluster rates is shown in figure \ref{fig:bariEffRate}. Despite the fact that the highest efficiency is reached with the STD mixture, the efficiency results in the range 90-79\% up to 800 Hz/cm$^2$ with the ECO2. Moreover, it is remarkable that in this case the plateau efficiency measured with the ECO3 gas mixture decreases more rapidly with respect to ECO2 and STD.

\begin{figure} [!h] 
    \centering
    \subfloat[\empty]{
        {\includegraphics[width=0.49\linewidth]{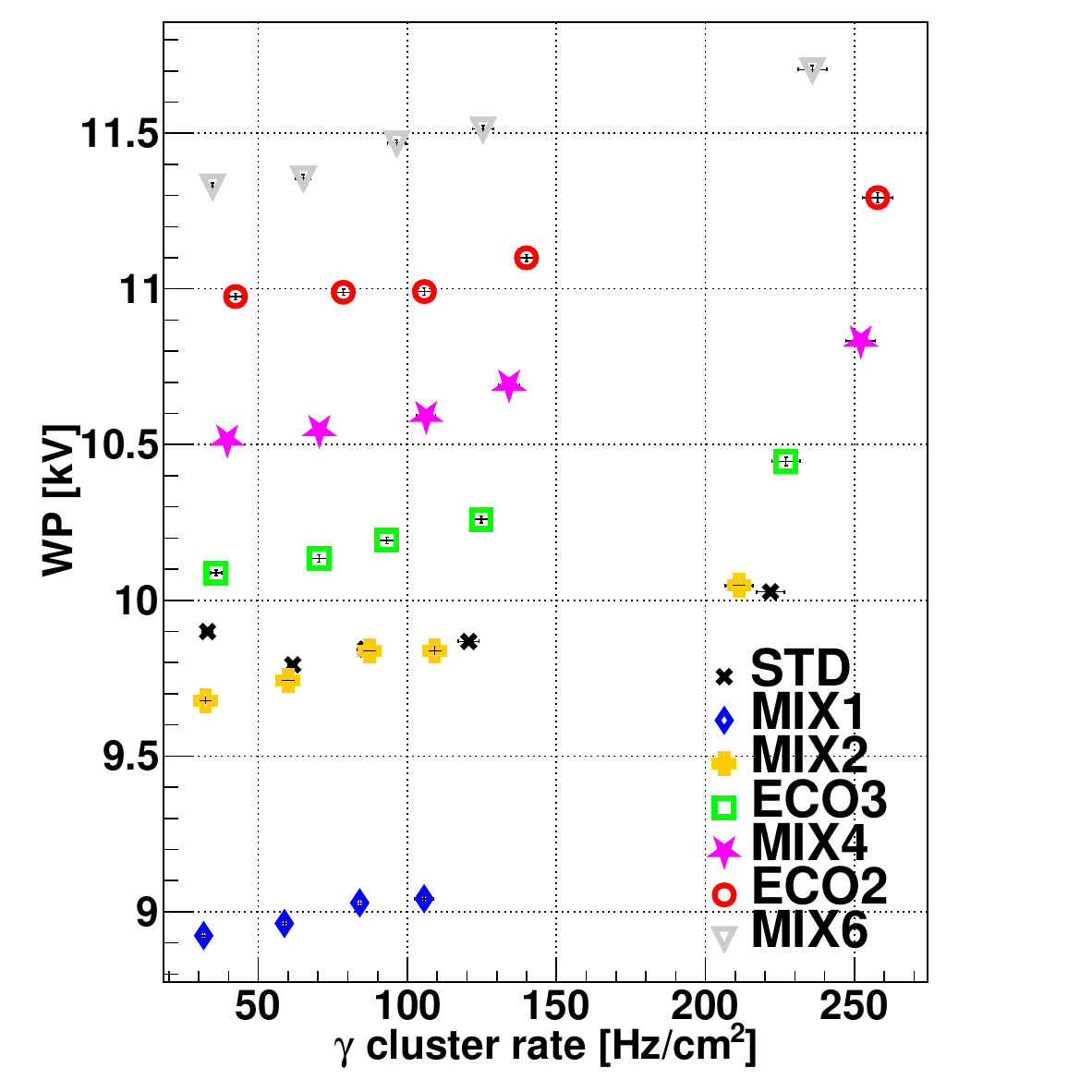}}
        \label{fig:AliceWPRate}
        }
    \subfloat[\empty]{
        {\includegraphics[width=0.49\linewidth]{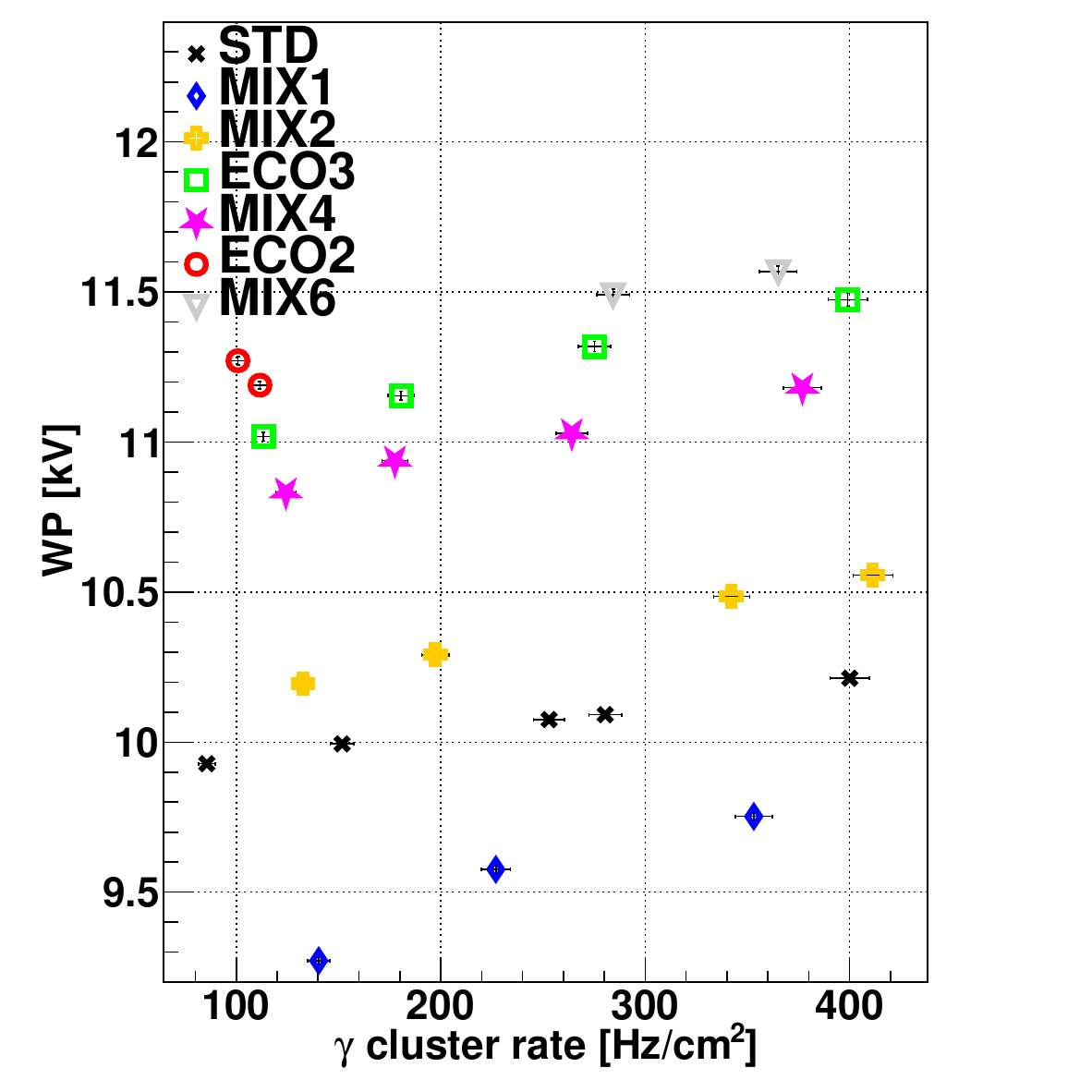}}
        \label{fig:epDtWPRate}
        }
    \caption{WP as a function of the $\gamma$ cluster rate measured at WP. Left panel: ALICE RPC. Right panel: EP-DT RPC}
    \label{fig:wp2mm}
\end{figure}

\begin{figure}[!h]
\centering
\includegraphics[width=0.49\linewidth]{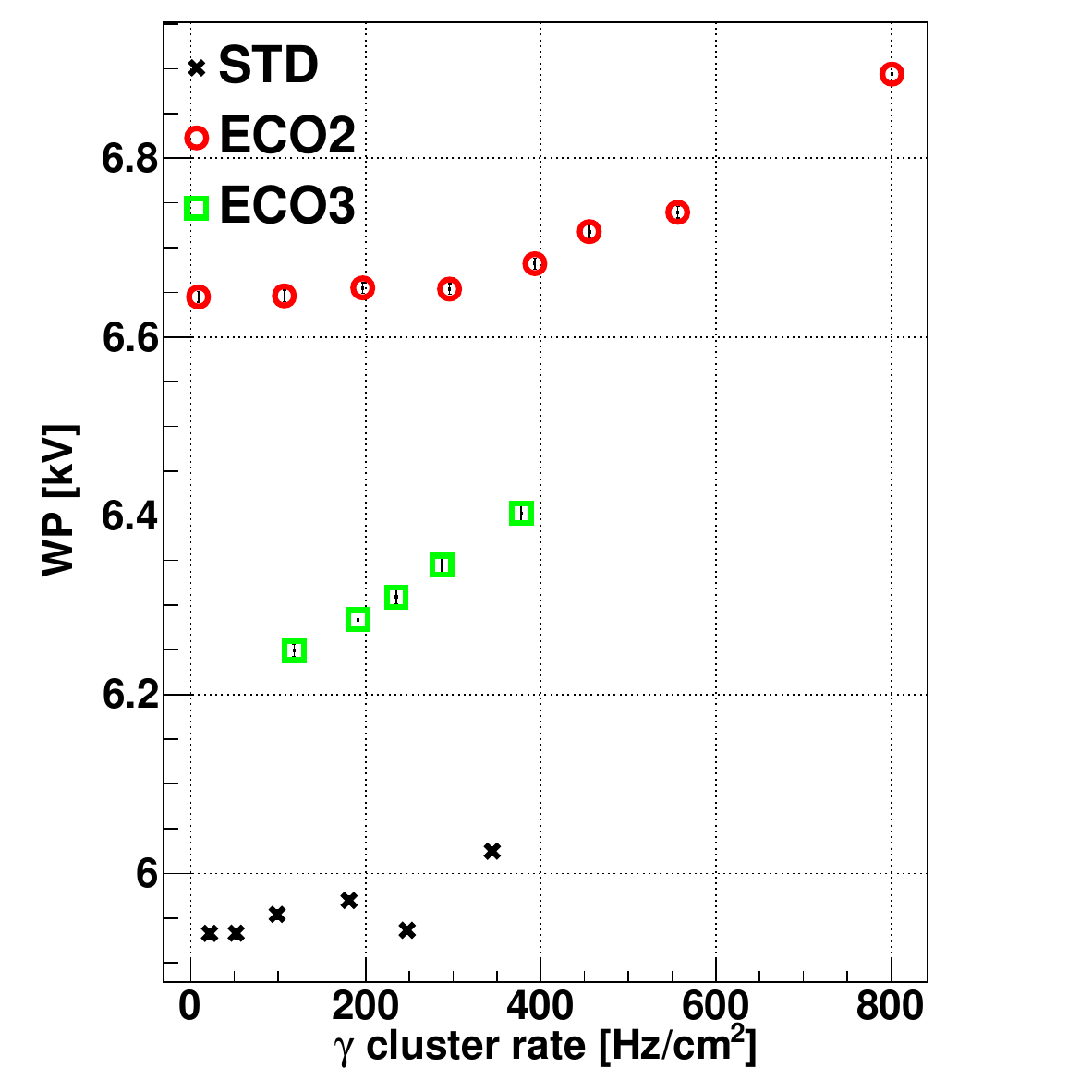}
\caption{WP for the BARI-1p0 detector as a function of the $\gamma$ cluster rate measured at WP}\label{fig:bariWPRate}
\end{figure}

The working points were calculated by interpolating the efficiency curves for each set of attenuation filters by using equation \ref{eq:fitEff}. Figures \ref{fig:wp2mm} and \ref{fig:bariWPRate} show the working points for the different $\gamma$ cluster rates measured with the ALICE, CERN EP-DT and BARI-1p0 RPCs. The ALICE and EP-DT detectors are characterized by similar working points and the value is shifted following the amount of CO$_{2}$ present in the mixtures. On the other hand, figure \ref{fig:bariWPRate} shows that the WP, for the BARI-1p0 RPC, characterized by a 1~mm gas gap and 1~mm electrodes, is shifted by just few hundred Volts, also when eco-friendly gas mixtures are used. The values, in this case, are shifted from 6.65 to 6.9~kV for ECO2 and from 6.3 to 6.45~kV for ECO3; a smaller shifts than for the detectors characterized by a 2~mm gas gap. The working point shift between mixtures is lower with respect to the ones calculated for ALICE and EP-DT RPCs due to the thinner gap used in the BARI-1p0 RPC. Though thinner gaps are characterized by a better timing performance, they have the disadvantage of a lower number of primary ion-electron clusters with respect to wider gaps filled with the same gas mixture. This consideration could explain why their maximum efficiency, as shown in figure \ref{fig:bariEffRate}, is lower with respect to detectors with thicker gaps. A mitigating solution is to have more than one thin gap in the same detector, which is the concept at the basis of multi-gap RPCs (in \cite{protoThesis}, the author has studied the behavior of multi-gap RPCs when operated with eco-friendly gas mixtures).

Thinner bakelite electrodes, on the other hand, have a beneficial effect in terms of rate capability, because the voltage drop at high rates across these planar electrodes, with reduced thickness, is lower with respect to thicker electrodes. From the practical point of view, one has to find the optimal compromise considering all the constructive parameters of a detector, given the requirements for the specific application. 

\subsubsection{Muon cluster size}

\begin{figure} [!h] 
    \centering
    \subfloat[\empty]{
        {\includegraphics[width=0.49\linewidth]{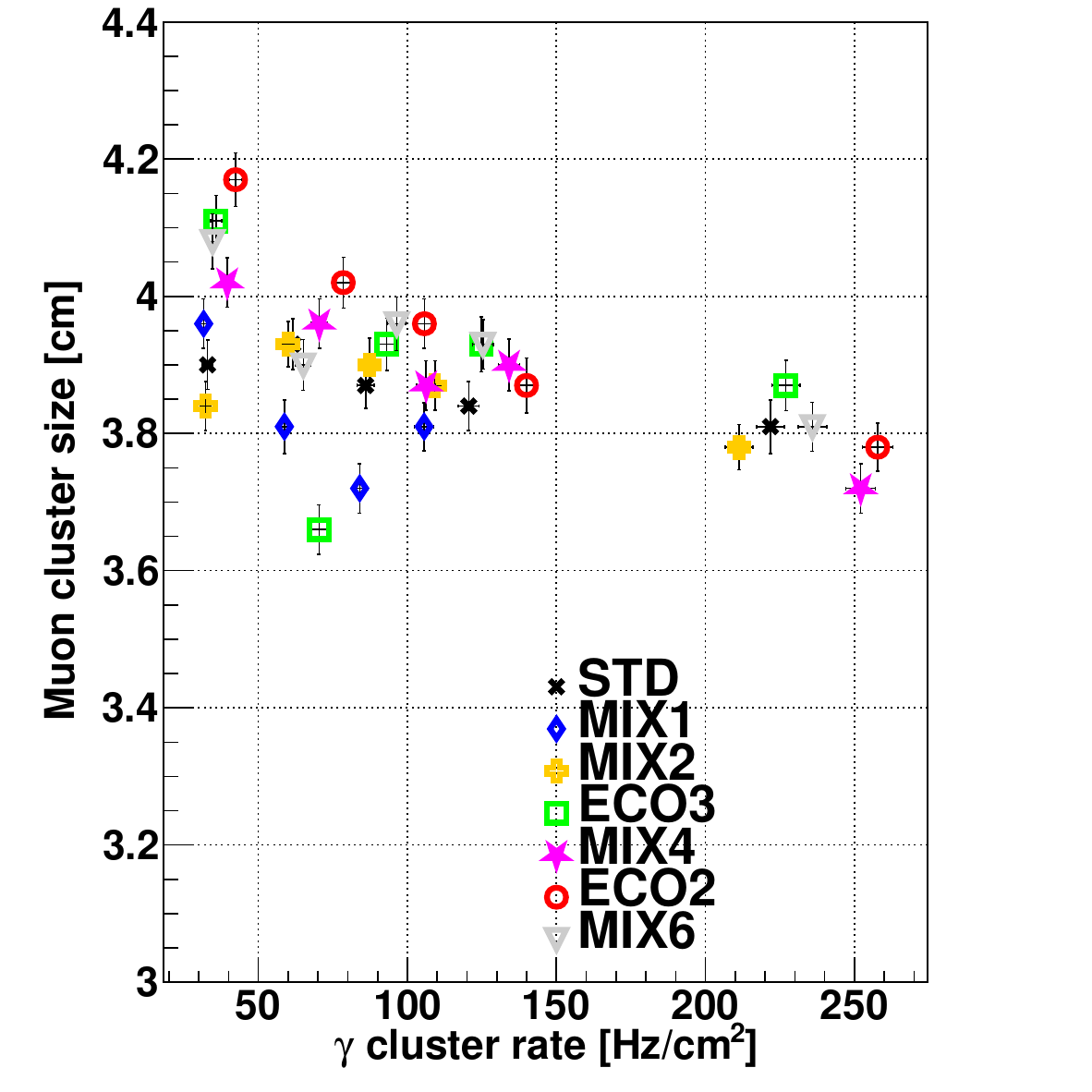}}
        \label{fig:AliceCLSRate}
        }
    \subfloat[\empty]{
        {\includegraphics[width=0.49\linewidth]{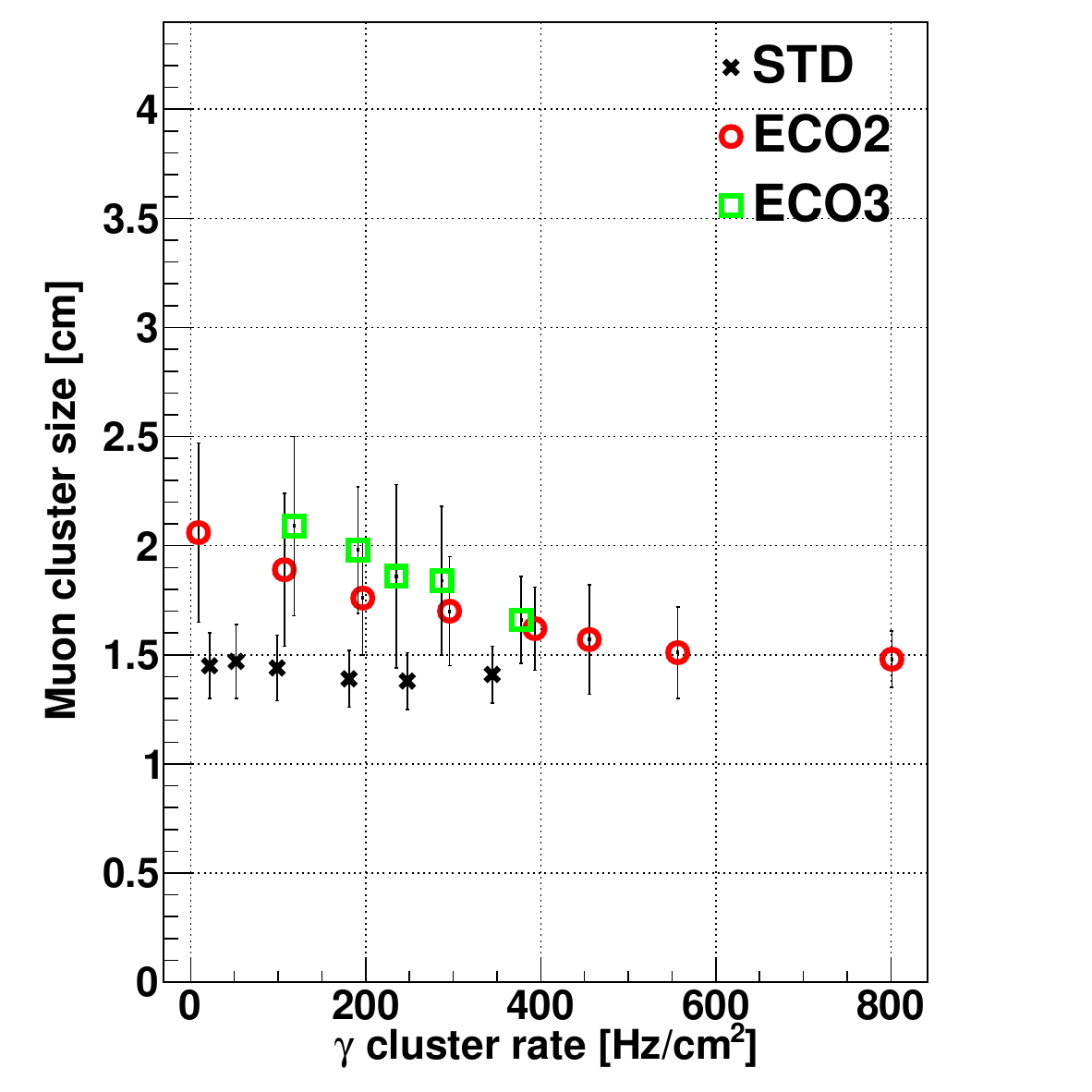}}
        \label{fig:epDtCLSRate}
        }
    \caption{Muon cluster size at WP as a function of $\gamma$ cluster rate measured at WP. Left panel: ALICE RPC. Right panel: BARI-1p0 RPC}
    \label{fig:clsrate}
\end{figure}

The muon cluster size quantifies the number of neighboring strips fired due to an avalanche produced by a muon. This value is is of significance because of its direct impact on the detector spatial resolution. In the following figures, the cluster size is expressed in~cm in order to enable the comparison between the different detectors with different strip pitch. Figure \ref{fig:clsrate} shows the muon cluster size values at WP as a function of the $\gamma$ cluster rate for the ALICE and BARI-1p0 detectors. In both RPCs, the values are slightly higher for the eco-friendly candidates at low $\gamma$ rates while at higher rates, the difference between the cluster sizes becomes less important. Moreover, one can observe that for all the tested gas mixtures, a decreasing trend with increasing irradiation is observed. 

\section{Preliminary aging studies}
\label{sec:agingStudies}

This section describes some preliminary results obtained from an aging test, covering the period from July 2022 to July 2023. First, a brief description of the general methodology used in the data-taking is provided; following this, a summary of the main results is presented.

\subsection{Methodology}
\label{subsub:methodology}

During the aging test, the detectors are flushed with the selected gas mixture, the high voltage is set to a fixed value and the stability of the absorbed current (measured by the high voltage power supply, with a precision of 0.1 $\mu$A) is monitored over time. The webdcs applies the correction for temperature/pressure changes, according to equation \ref{eq:corr}, as explained in section \ref{sec:intro}, in order to maintain a constant HV$_{eff}$ on the detectors. 

The values of current, applied and effective high voltage are saved every 30 seconds for data analysis. Moreover, once a week, the $^{137}$Cs source is fully shielded (source-off) and a measurement of the absorbed current without $\gamma$ irradiation is performed. This current is familiarly called "dark current" and it is an important parameter to monitor throughout the aging test, since its increase could be a sign of detector aging.

\begin{figure} []
    \centering
    \subfloat{{\includegraphics[width=0.49\linewidth]{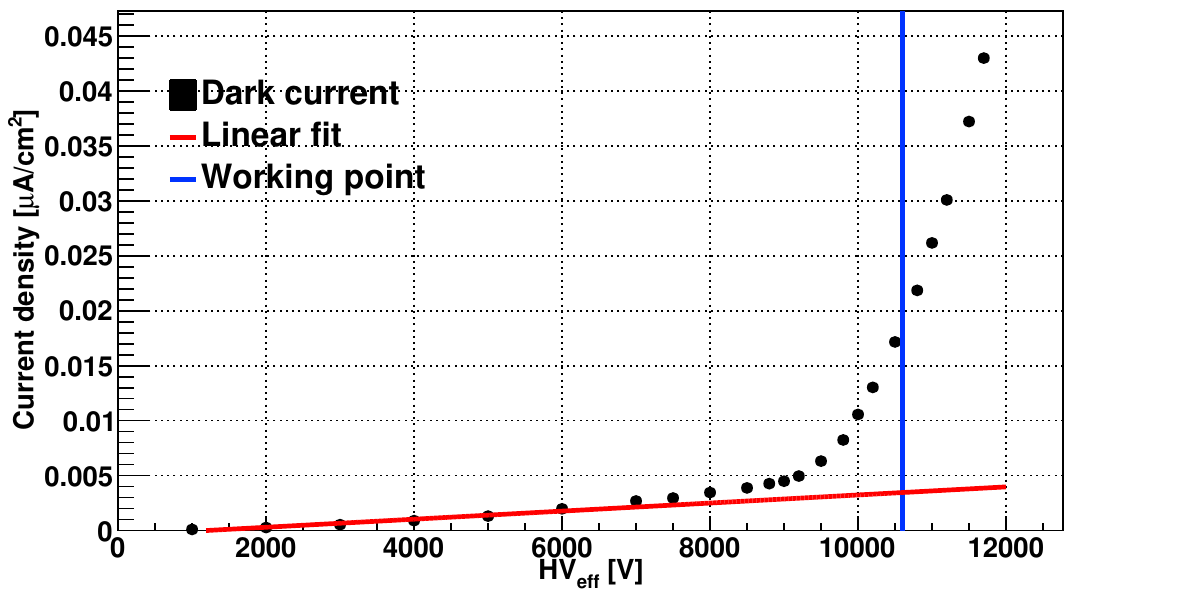}}}
    \subfloat{{\includegraphics[width=0.49\linewidth]{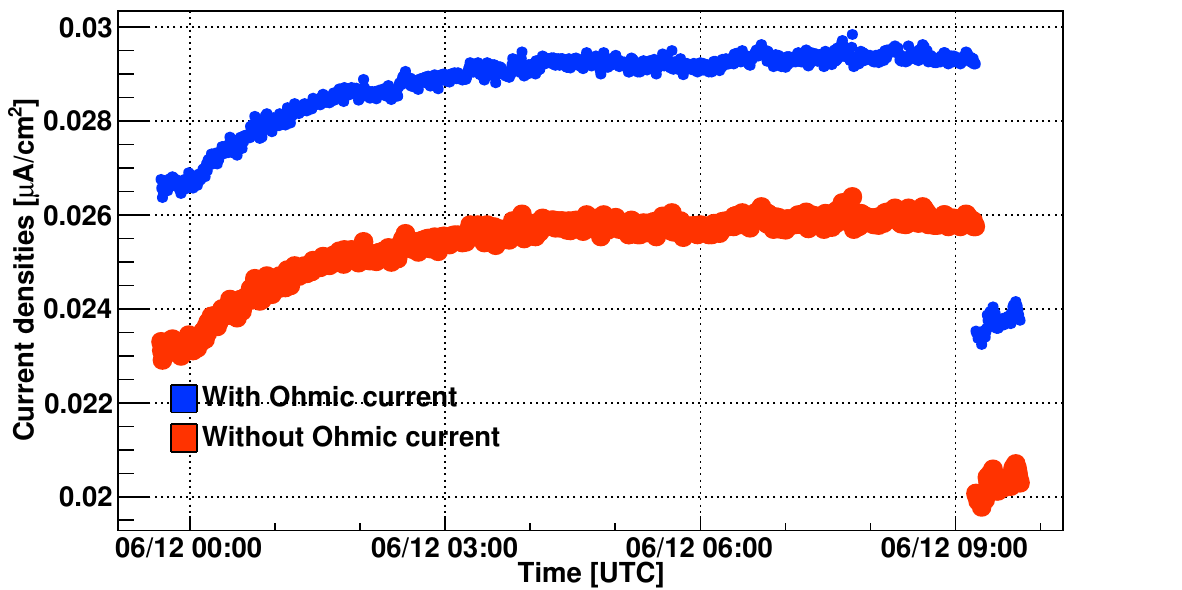}}}
    \caption{Left panel: example of the dark current density as a function of HV$_{eff}$ for the EP-DT detector. The Ohmic part of the dark current is clearly visible as well as the linear interpolation to obtain the Ohmic dark current at the irradiation voltage (represented by the intersection between the vertical blue line and the red straight line). Right panel: example of current density under irradiation with and without the Ohmic part of the dark current (EP-DT detector). The discrete step observed towards the end of the period corresponds to a change in the irradiation conditions, which causes to a current reduction}
    \label{fig:dark1}
\end{figure}

In order to numerically quantify the progress of the aging, one can use the \textit{integrated charge density}, defined as the integral over time of the current density passing through the detector and measured in mC/cm$^{2}$. By looking at the left panel of figure \ref{fig:dark1}, one can see an example of the absorbed dark current density as a function of HV$_{eff}$ (I(HV) curve) for the EP-DT detector when flushed with the ECO2 gas mixture; it is possible to observe that, even for voltages well below the threshold for avalanche multiplication processes (7/8~kV with this mixture and for a 2~mm gap, as it is the case for EP-DT detector), a non-zero current is flowing. This is defined as the Ohmic component of the dark current and it is, in principle, not flowing through the gas but rather through some other conductive paths in the detector. Most likely, this Ohmic component is not relevant for aging processes, since it is not related to discharge processes happening in the gas, which may lead to the dissociation of HFO and the production of harmful pollutants. Source-off current density vs HV curves are measured weekly and this allows one to also monitor the Ohmic component of the dark current.

Since the irradiation is carried out at fixed HV$_{eff}$ (10.6~kV for the 2~mm gaps and 8.8-9~kV for the 1.6~mm one, in the case of the ECO2 gas mixture, as it will be better explained in section \ref{subsub:agingResults}), it is useful to estimate the Ohmic part of the dark current at said voltage. To this aim, a linear interpolation of the current density vs HV$_{eff}$ curve is carried out between 0 and 5~kV, for the 2~mm gaps and between 0 and 4~kV for the 1.6~mm one. The straight line is then extrapolated to the irradiation voltage, providing the required estimation of the Ohmic dark current density. This is then subtracted from the total current density measured, obtaining the current density flowing through the gas and related to the $\gamma$ irradiation. Since a weekly dark current measurement is performed, one can subtract its Ohmic component from the measured current density for the whole irradiation period (i.e. for each irradiation period we subtract its closest (in time) estimation of Ohmic dark current). An example of this procedure is shown in the right panel of figure \ref{fig:dark1}, where the current density absorbed under irradiation is shown for the EP-DT detector. The blue markers represent the current density measured by the high voltage module, the ones in red show the current when the Ohmic component was subtracted. The current density to which its Ohmic component was subtracted is used to compute the integrated charge density.

\subsection{Main results from aging studies}
\label{subsub:agingResults}

During the irradiation studies carried out between July 2022 and July 2023, the detectors have been flushed with the ECO2 gas mixture and, for the most of these studies, the source attenuation filter was set to a value of~2.2. As anticipated, the HV$_{eff}$ chosen for the irradiation corresponds to 10.6~kV (for the 2~mm gas gap detectors) and 8.8/9~kV (for the 1.6~mm SHiP RPC, detector fully characterized with beam whose results have not been shown in this paper for the sake of avoiding repetition). Note that the BARI-1p0 detector was not included in the aging studies since it experienced a large current increase following the beam test campaigns and it was removed from the setup. With this HV$_{eff}$ the detectors are not fully efficient. The reason why this voltage was chosen is to limit the currents absorbed by the detectors (indeed, by looking at figure \ref{fig:atlasOnHV} the current absorbed with a background of 500~Hz/cm$^{2}$ with ECO2 is $\approx$twice as much as with respect to the standard gas mixture). Moreover, in the LHC experiments the detectors are not kept with high voltage on at all times, hence aging studies with too large currents drawn continuously for too long would not be representative of real life conditions, possibly leading to the appearance of artifacts which would not be observed when operating the detectors. 

\begin{figure} [H]
    \centering
    \subfloat{{\includegraphics[width=0.49\linewidth]{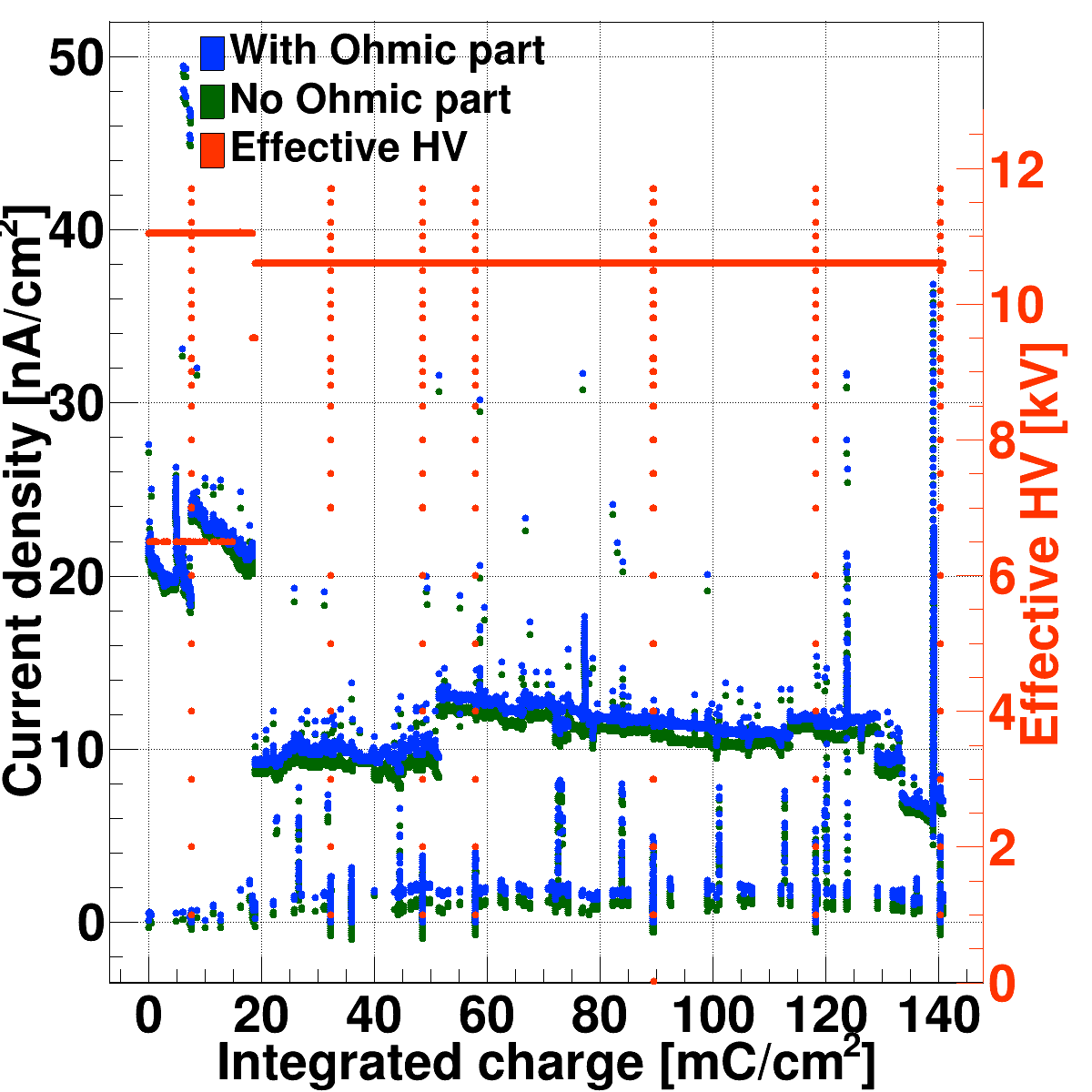}}}
    \subfloat{{\includegraphics[width=0.49\linewidth]{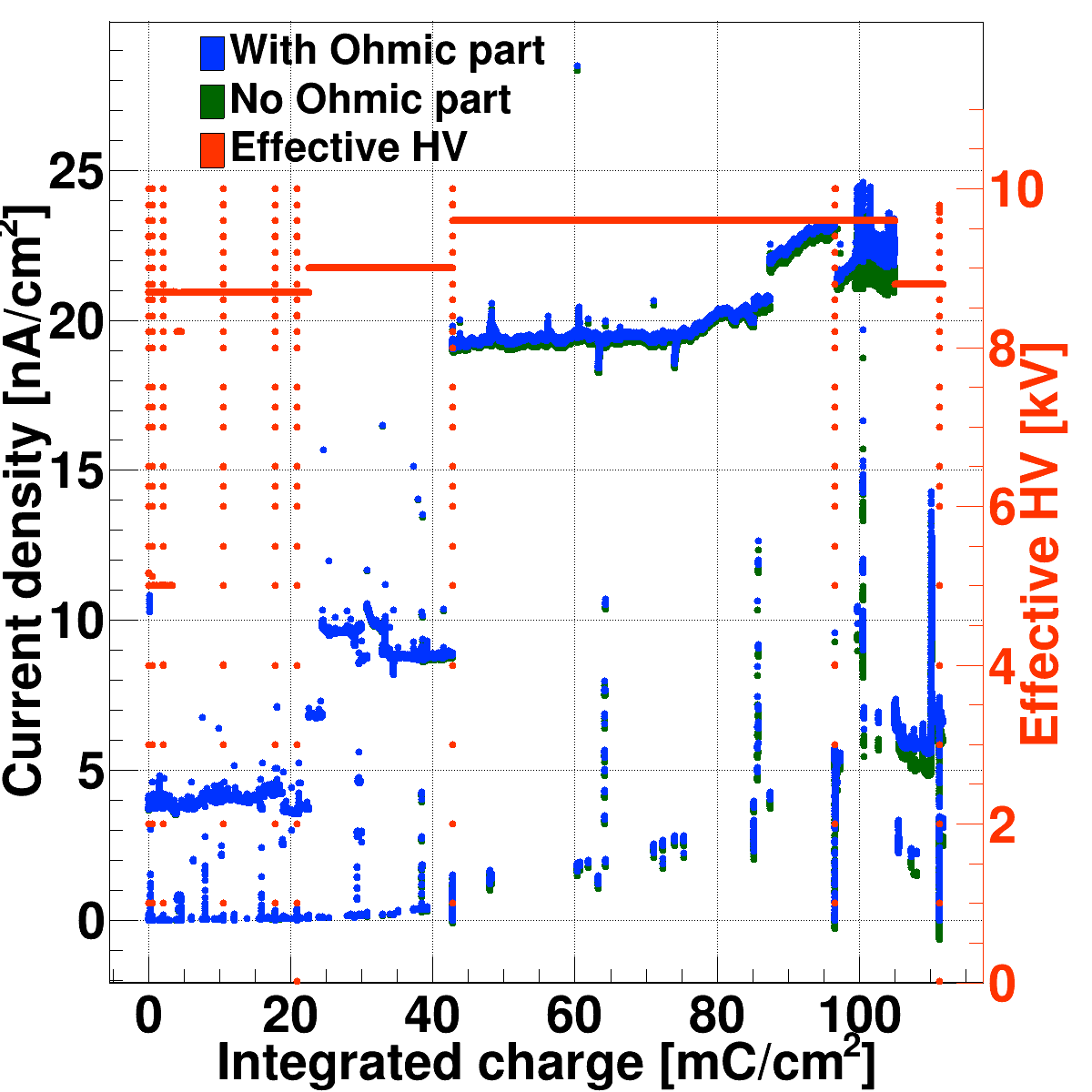}}}
    \caption{Absorbed current density (with and without the Ohmic dark current) and HV$_{eff}$ during an $\approx$ one year exposure to the GIF++ $^{137}$Cs source as a function of the integrated charge density. Left panel: CMS RE11 TN gap. Right panel: SHiP detector}
    \label{fig:agingAllECO2}
\end{figure}

Figure \ref{fig:agingAllECO2} summarizes the trend of the absorbed current density and HV$_{eff}$ applied to the RPCs during the whole aging campaign. All the quantities in the figure are shown as a function of the integrated charge density, since the irradiation campaign is sometimes interrupted and this would leave empty gaps in the chart. The left panel of figure \ref{fig:agingAllECO2} shows the results for the CMS RE11 TN gap while the right panel for the SHiP detector.

The HV$_{eff}$ is shown in red and, as expected, it is constant throughout the whole irradiation campaign. The vertical dotted lines correspond to the weekly source-off dark current vs HV$_{eff}$ scans mentioned in section \ref{subsub:methodology} (Those reported in figure \ref{fig:agingAllECO2} do not correspond to all the scans taken throughout the aging campaign but it sometimes happened that the data of a dark current were saved together with those of the aging studies and they are reported in this figure).

For what concerns the absorbed current density, the figure shows both the total one (in blue), as well as the one with the subtraction of the Ohmic part of the dark current (in green). As it was explained in section \ref{subsub:methodology}, the latter is used to compute the integrated charge density. The current values shown in the figure are independent of the source status and, during the irradiation period, it sometimes occurs that the source is fully shielded due to other users' requests or other interventions to the facility (beside the weekly source-off day mentioned in \ref{subsub:methodology}). This observation explains the presence of two distinct populations in figure \ref{fig:agingAllECO2}. In the left panel of the figure, for example, the values around 0-1~nA/cm$^{2}$ correspond to the current density absorbed when the source is fully shielded and the one around 10~nA/cm$^{2}$ is the current absorbed under irradiation. The portions of the trend where the current varies rapidly correspond to the source-off scans described earlier (indeed, they are always accompanied by changing HV$_{eff}$ values).

In the case of the SHiP detector (right panel of figure \ref{fig:agingAllECO2}), the HV$_{eff}$ was increased in steps from 8.8~kV, corresponding to 50\% efficiency, up to 9.8~kV, corresponding to plateau efficiency, in order to study the evolution of the current accordingly. It is possible to see that, towards $\approx$80~mC/cm$^{2}$, the absorbed current starts to fluctuate in a more pronounced way. To investigate this effect, the HV$_{eff}$ of the SHiP RPC was reduced and the current absorbed with this lower value is being closely monitored. Indeed, as reported in \cite{abbrescia}, keeping the detectors with a lower than nominal applied high voltage has been proven to somewhat reduce the current drawn by the detectors (a possible explanation for this might be the \textit{burn} of small imperfections, created due to chemical interactions, on the side of the bakelite electrode exposed to the gas). 

\begin{figure} []
    \centering
    \subfloat{{\includegraphics[width=0.49\linewidth]{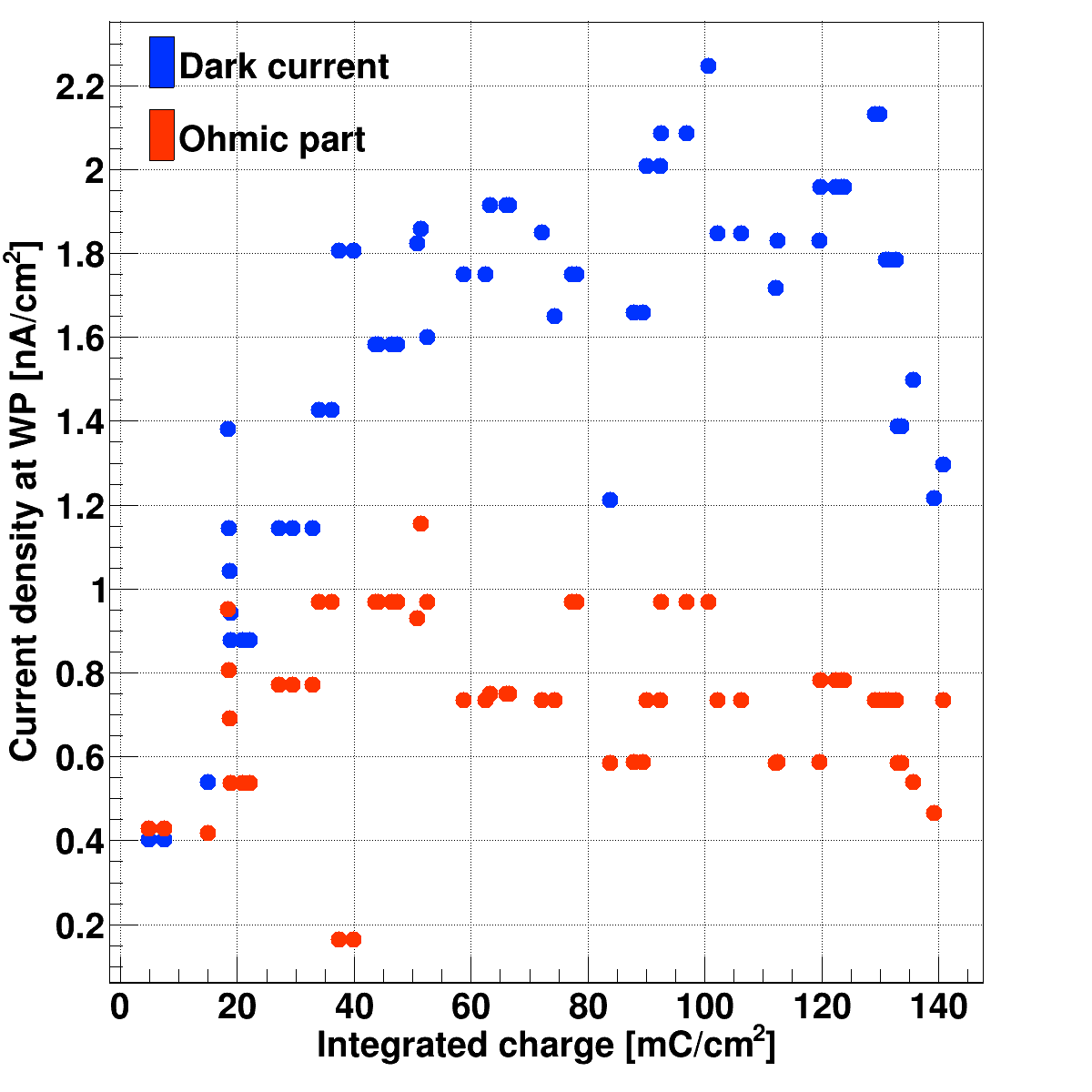}}}
    \subfloat{{\includegraphics[width=0.49\linewidth]{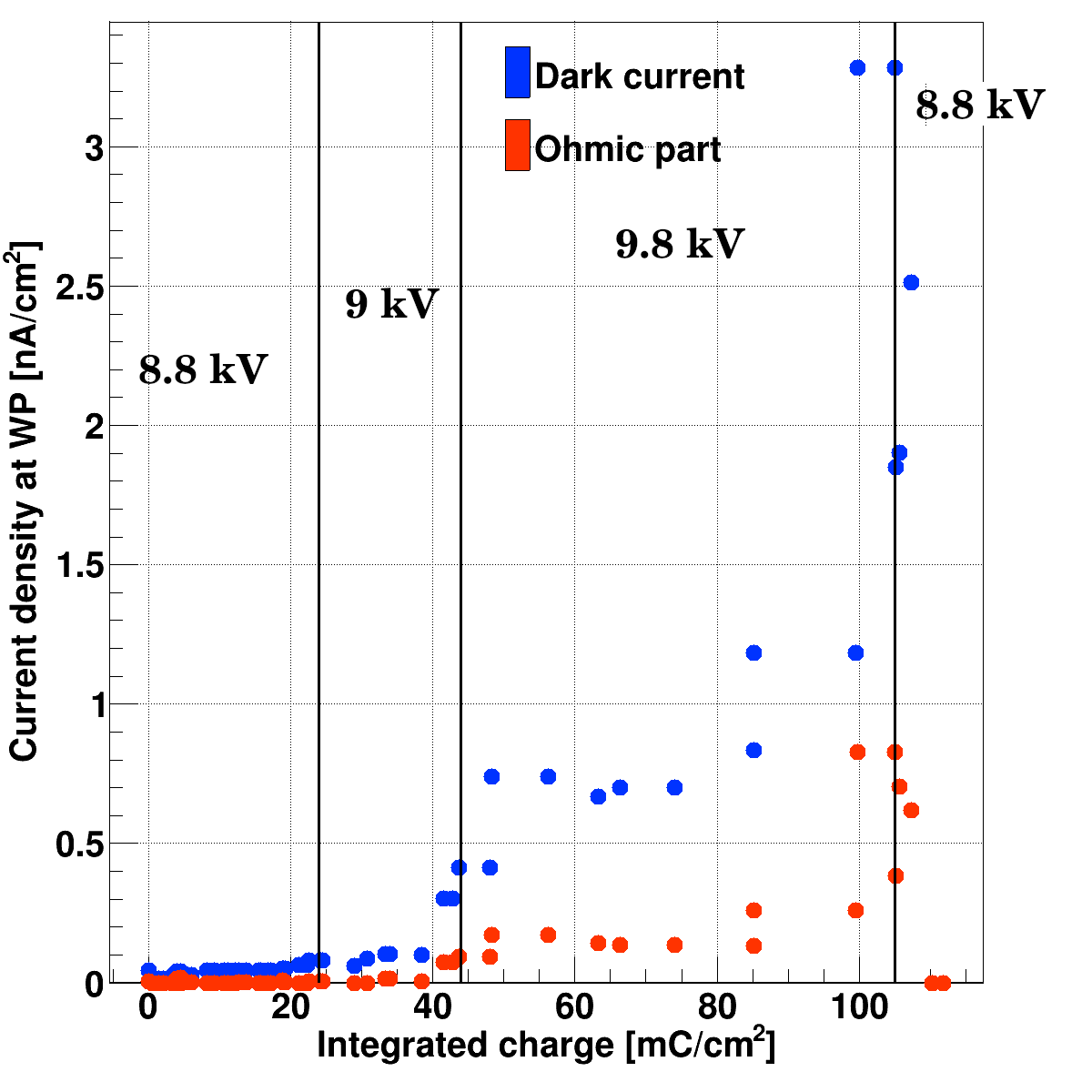}}}
    \caption{Dark current density (total and extrapolated Ohmic part) at the irradiation voltage (10.6 and 8.8 up to 9.8~kV for the 2 and 1.6~mm gaps respectively) as a function of the integrated charge density following an exposure of around one year to the GIF++ $^{137}$Cs source. Left panel: CMS RE11 TN gap. Right panel: SHiP detector (the different HV$_{eff}$values are also reported on the chart for this detector)}
    \label{fig:darkCurrentAll}
\end{figure}

It is also useful to monitor the evolution of the dark current (both its Ohmic as well as the total components). Indeed, as anticipated, an increase of absorbed dark current could be a sign of potential detector aging. Figure \ref{fig:darkCurrentAll} summarizes this trend, by showing both components of the dark current at the irradiation voltage (10.6~kV for the 2~mm detectors and between 8.8 and 9.8~kV for the 1.6~mm gap), as a function of the integrated charge density for the CMS RE11 TN gap (left panel) and the SHiP detector (right panel). 

In figure \ref{fig:darkCurrentAll}, one can see that, sometimes, the same current density value is reported for different integrated charge densities. This is due to the fact that a single source off current scan is performed per week but multiple irradiation scans could be started and stopped during the same week, hence the same value of dark current density is used for multiple irradiation scans. The discrete step that can be seen in the right panel of figure \ref{fig:darkCurrentAll} at $\approx$40~mC/cm$^{2}$ corresponds to the fact that the irradiation voltage for the SHiP detector was increased, hence the higher values. The last few points of the same chart refer to the abnormal increase in absorbed current density already reported in figure \ref{fig:agingAllECO2}.

Also, from the left panel of figure \ref{fig:darkCurrentAll}, it is possible to see that the Ohmic component of the dark current shows an increasing trend at the start of the irradiation, while it reaches a more stable behavior for higher values of integrated charge density. For what concerns the total dark current density, it shows a more uniform increasing trend during the whole irradiation campaign. 

\begin{figure} []
    \centering
    \subfloat{{\includegraphics[width=0.49\linewidth]{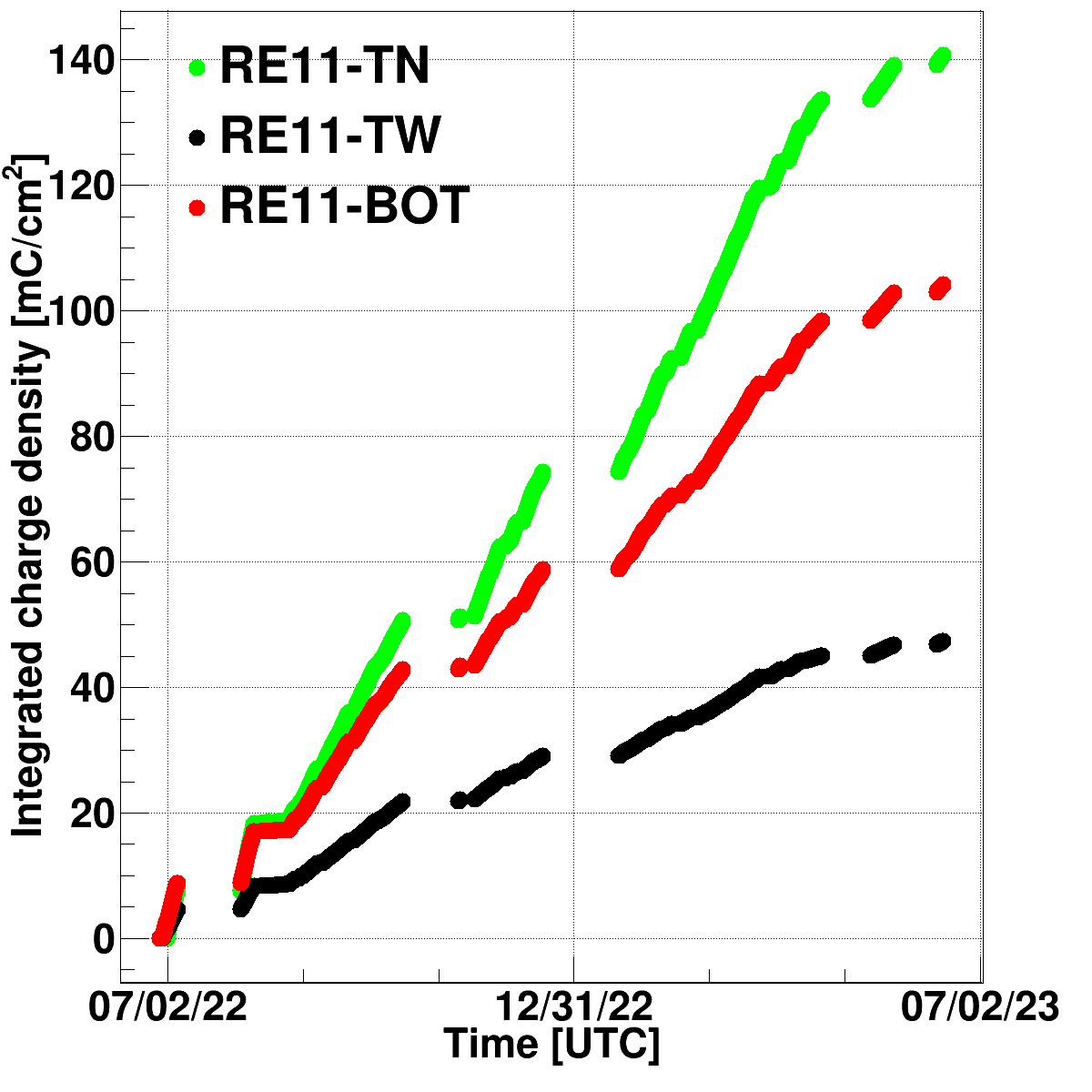}}}
    \subfloat{{\includegraphics[width=0.49\linewidth]{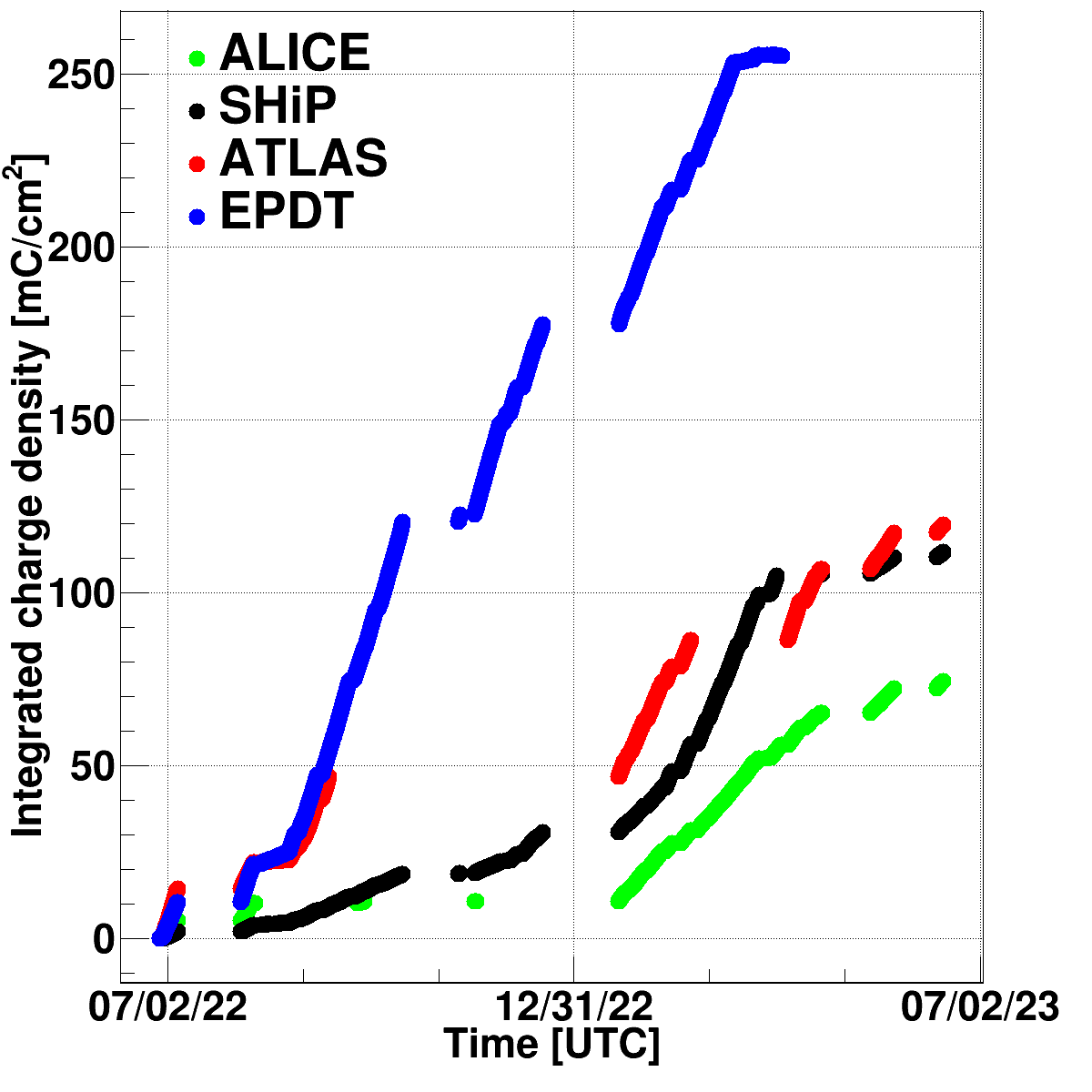}}}
    \caption{Accumulated charge density as a function of time for all the RPC ECOgas@GIF++ collaboration detectors. Left panel: CMS RE11 RPC (three gaps). Right panel: ALICE, ATLAS, EP-DT and SHiP RPCs}
    \label{fig:intCharge}
\end{figure}

The integrated charge density during around one year of exposure to the GIF++ $^{137}$Cs source, is shown in figure \ref{fig:intCharge}. The left panel shows the results for the three gaps of the CMS RE11 RPC while the right panel refers to the other detectors. The fact that the integrated charge density is not exactly the same across the detectors, can be explained by considering that the irradiation voltage chosen does not exactly correspond to the same efficiency value.

As it was stated earlier, the results obtained so far are preliminary and, for the moment, a clear behavior cannot be pointed out. The behavior of some specific detectors (which showed a more significant increase of the total dark current density than others) needs to be closely monitored in time and, in order to shed some light on this, the RPC ECOgas@GIF++ collaboration is planning to start the monitoring of other parameters, such as the presence of possible current leaks on the mechanical frame and the production of fluorinated impurities in the exiting gas mixture. Moreover, one also needs to monitor the detectors  performance in terms of response to cosmic/beam muons, with time. This has been done in July 2023, when another beam test campaign was carried out and the data gathered is currently being analyzed. In this way, one will be able to estimate the performance evolution and have a first insight on the real aging observed on the detectors.

\section{Conclusions}
\label{sec:conclusion}

This paper explored some of the most recent activities of the RPC ECOgas@GIF++ collaboration. These have been focused on performance and aging studies on RPC detectors operated with different eco-friendly gas mixtures, where C$_{2}$H$_{2}$F$_{4}$ has been replaced using mixtures with various concentrations of HFO and CO$_{2}$. 

During the beam tests, it was observed that the plateau efficiency reached without irradiation increases at increasing HFO concentrations and so does the detector WP. It was observed that this shift is around 1~kV for every 10\% HFO added to the gas mixture.

The average value of signal charge, for all the HFO-based gas mixtures, is generally larger, with respect to the standard gas mixture and also a higher fraction of events with large charge content is observed. Both values tends to decrease if the HFO concentration increases; reaching, at the detector WP, similar values to the standard gas mixture. It was nevertheless observed that the useful operating region (i.e. the high voltage range where the efficiency is above 95\% and the large signal contamination is below 5\%) is reduced for the eco-friendly alternatives (since the number of events with large signals increases more sharply with the voltage, if compared to the standard gas mixture).

For what concerns the RPC response under $\gamma$ irradiation, different observations can be made: first of all, the efficiency curves shift to higher voltages (the same can be said also for the detector WP) if the background level increases; secondly, the plateau efficiency decreases. These effects can be partly explained by considering that when the detectors are exposed to an intense $\gamma$ background, the absorbed current increases and, circulating through the resistive bakelite electrodes, this leads to a voltage drop across the electrodes themselves, leading to a reduction of the voltage applied to the gas, leading to a lower gain and lower efficiency. It was observed that the maximum efficiency reduction is $\approx$ 1-2 \% (between source off and the highest irradiation condition) for the standard gas mixture while it ranges from 8 down to 4 \% for the HFO-based gas mixtures (the effect is less pronounced if more HFO is added to the mixture).

For what concerns the preliminary results obtained from the aging studies, an irradiation campaign with the ECO2 (35/60~HFO/CO$_{2}$) was started in July 2022. The stability of the absorbed current (both with and without irradiation) was monitored for around one year now. It was observed that the current under irradiation is quite stable over time. The Ohmic component of the dark current also appears quite stable over time while the trend of the total dark current is more subject to fluctuations. These effects are under investigation at the moment but what would be of the utmost importance in the future is a continuous monitor of all the detector performance (i.e. efficiency, prompt charge, pulse spectrum etc.). A beam test campaign has been carried out in July 2023, to perform a first comparison with the previous data; the analysis is still ongoing and, soon, this comparison will be made.

All in all, the efforts of the RPC ECOgas@GIF++ collaboration have led to some breakthrough in the search for eco-friendly alternative gas mixtures. The ongoing aging campaign, complemented by periodic beam test studies, will help to shed some light on the long-term behavior of RPC detectors operated with eco-friendly alternatives studies in this manuscript.

\section*{Declarations}

\paragraph*{Funding} This work is part of the AIDA Innova project, which received funding from the European Union’s Horizon 2020 Research and Innovation programme under GA n. 101004761. In particular, this work is part of the Working Package 7.2 (Multigap RPCs (MRPCs) for fast timing and Eco-friendly gas mixtures for RPCs)

\paragraph*{Data availability statement} The data shown in this article can be made available by the corresponding author(s) upon request

\bibliography{sn-bibliography}

\end{document}